\documentclass[twocolumn,apjfonts]{aastex631}

\usepackage{xcolor}
\usepackage{textgreek}
\usepackage[utf8]{inputenc}
\usepackage[english]{babel}

\usepackage{multirow}

\usepackage{hyperref}
\hypersetup{
    unicode, 
    colorlinks=true,
    linkcolor=linkcolor,
    citecolor=linkcolor,
    filecolor=linkcolor,
    urlcolor=linkcolor,
}
\usepackage{color,colortbl}
\definecolor{linkcolor}{rgb}{0.0,0.3,0.5}

\DeclareGraphicsExtensions{.bmp,.png,.jpg,.pdf}
\usepackage{verbatim}
\usepackage[normalem]{ulem}
\usepackage{orcidlink}
\usepackage{soul}

\usepackage{booktabs}
\usepackage{multirow}

\usepackage{xspace}
\usepackage{xcolor}
\newcommand{\new}[1]{\textcolor{black}{#1}}
\newcommand{\revthree}[1]{\textcolor{black}{#1}}
\newcommand{\revtwo}[1]{\textcolor{black}{#1}}
\newcommand{\rev}[1]{\textcolor{black}{#1}}

\newcommand{\Swift}[0]{\textit{Swift}\xspace}
\newcommand{\XMM}[0]{\textit{XMM-Newton}\xspace}
\newcommand{\eROSITA}[0]{\textit{eROSITA}\xspace}
\newcommand{\Chandra}[0]{\textit{Chandra}\xspace}

\newcommand{\Object}[0]{\texttt{Object}}
\newcommand{\DiaObject}[0]{\texttt{DiaObject}}
\newcommand{\Source}[0]{\texttt{Source}}
\newcommand{\diffIm}[0]{\texttt{difference\_image}}
\newcommand{\visitIm}[0]{\texttt{visit\_image}}
\newcommand{\deepCoadd}[0]{\texttt{deep\_coadd}}
\newcommand{\DiaSource}[0]{\texttt{DiaSource}}
\newcommand{\ForcedSource}[0]{\texttt{ForcedSource}}
\newcommand{\ForcedSourceOnDiaObject}[0]{\texttt{ForcedSourceOnDiaObject}}

\begin{document}
\title{Optical Counterparts to X-ray sources in LSST DP1}

\author[0000-0001-5538-0395]{Yuankun (David) Wang}
\email{ykwang@uw.edu}
\affiliation{DIRAC Institute, Department of Astronomy, University of Washington, 3910 15th Avenue NE, Seattle, WA 98195, USA}

\author[0000-0001-8018-5348]{Eric C.\  Bellm}
\affiliation{DIRAC Institute, Department of Astronomy, University of Washington, 3910 15th Avenue NE, Seattle, WA 98195, USA}

\author[0000-0003-3318-0223]{Robert I.\ Hynes}
\affiliation{Department of Physics and Astronomy, Louisiana State University, Baton Rouge, LA 70803-4001, USA}

\author[0000-0002-9547-8677]{Yue Zhao}
\affiliation{School of Physics \& Astronomy, University of Southampton, SO17 1BJ, UK}

\author[0000-0003-3105-2615]{Poshak Gandhi}
\affiliation{School of Physics \& Astronomy, University of Southampton, SO17 1BJ, UK}

\author[0000-0002-9396-7215]{Liliana Rivera Sandoval}
\affiliation{Dept. of Physics and Astronomy, University of Texas Rio Grande Valley, Brownsville, TX 78520, USA}

\author[0009-0007-9870-9032]{Sandro Campos}
\affiliation{The McWilliams Center for Cosmology \& Astrophysics, Department of Physics, Carnegie Mellon University, Pittsburgh, PA 15213, USA}

\author[0000-0003-3287-5250]{Neven Caplar}
\affiliation{DIRAC Institute, Department of Astronomy, University of Washington, 3910 15th Avenue NE, Seattle, WA 98195, USA}

\author[0000-0002-1074-2900]{Melissa DeLucchi}
\affiliation{The McWilliams Center for Cosmology \& Astrophysics, Department of Physics, Carnegie Mellon University, Pittsburgh, PA 15213, USA}

\author[0000-0001-7179-7406]{Konstantin Malanchev}
\affiliation{The McWilliams Center for Cosmology \& Astrophysics, Department of Physics, Carnegie Mellon University, Pittsburgh, PA 15213, USA}

\author[0000-0002-1174-2873]{Amy Secunda}
\affiliation{Center for Computational Astrophysics, Flatiron Institute, 162 Fifth Avenue, New York, NY 10010, USA}

\author[0000-0001-6320-2230]{Tobin M. Wainer}
\affiliation{DIRAC Institute, Department of Astronomy, University of Washington, 3910 15th Avenue NE, Seattle, WA 98195, USA}

\begin{abstract}
    We present a crossmatch between a combined catalog of X-ray sources and the Vera C. Rubin Observatory Data Preview 1 (DP1) to identify optical counterparts. The six fields targeted as part of DP1 include the Extended Chandra Deep Field South (E-CDF-S), the Euclid Deep Field South (EDF-S), the Fornax Dwarf Spheroidal Galaxy (Fornax dSph), 47 Tucanae (47 Tuc) and science validation fields with low galactic and ecliptic latitude (SV\_95\_-25 and SV\_38\_7, respectively). We find matches to 2314 of 3830 X-ray sources. We also compare our crossmatch to DP1 in the E-CDF-S field to previous efforts to identify optical counterparts. The probability of a chance coincidence match varies across each DP1 field, with overall high reliability in the E-CDF-S field, and lower proportion of high-reliability matches in the other fields. The majority of previously known sources that we detect are, unsurprisingly, active galaxies. We plot the X-ray-to-optical flux ratio against optical magnitude and color in an effort to identify Galactic accreting compact objects using a {\em Gaia} color threshold transformed to LSST $g$--$i$, but do not find any strong candidates in these primarily extragalactic counterparts. The DP1 dataset contains high-cadence photometry collected over a number of nights. We calculate the Stetson \( J \) variability index for each object under the hypothesis that X-ray counterparts tend to exhibit higher optical variability; however, the evidence is inconclusive whether our sample is more variable over DP1 timescales when compared to field objects.
\end{abstract}




\section{Introduction}
\label{sec:intro}
The X-ray sky provides a window into high-energy processes in extreme environments across the universe. X-ray surveys offer an important way to discover a range of interesting and rare objects such as low mass X-ray binaries (LMXBs), \rev{high mass X-ray binaries (HMXBs)}, binary and isolated millisecond pulsars (MSPs), cataclysmic variables (CVs), as well as other objects like \rev{active stars, galaxies,} AGN\new{, and galaxy clusters}. Over the last 25 years, missions such as \Chandra, \XMM, \Swift, and \eROSITA\ have revealed about a million sources, a small fraction of which have been classified with the aid of multiwavelength observations, while the majority remain unidentified.

Since the localization of X-ray sources is generally less precise than that of optical catalogs, the higher density of the latter often results in multiple optical sources falling within a single X-ray error circle. As a result, a variety of techniques have been developed to evaluate the strength of each potential match. One common technique is known as the likelihood ratio (LR) method, which estimates the probability that a given counterpart is the true match given the local stellar densities and distribution \citep{Richter:1975:LR, Sutherland:1992:LikelihoodRatio}. This is among the most widely adopted techniques, such as in the crossmatch between X-ray and optical sources in the ROSAT Bright source catalog and the USNO A-2 optical catalog \citep{Rutledge:00:RASSXID}, or the Chandra Galactic Bulge Survey \citep{Weveres:2016:GBS_xmatch}. More recently, Bayesian methods that incorporate priors and probabilities for probabilistic cross-identification have been developed \citep{Budavari:08:BayesianCrossmatch, Budavari:16:Probabilistic-C, Shi:19:ILPCrossmatch} and these methods have been also applied to the ROSAT all sky survey \citep{Haakonsen:09:XIDII}. Some of these techniques, such as {\sc nway} \citep{Salvato:2018:NWAY} can evaluate multiple candidates or null matches across multiple catalogs jointly. In the past few years, \cite{Bykov:2022:eROSITA_Lockman_ML} used machine learning techniques, training a neural network to perform an optical crossmatch of \eROSITA\ sources in the Lockman Hole.

All of these techniques assume knowledge of the photometric priors, which is typically derived from known matched samples. However, the training and validation datasets for these models must be based on unambiguous matches \citep{Salvato:2018:NWAY}. In addition, many published crossmatches focus on a small area of the sky, such as the Lockman Hole Field, which display advantageous qualities, such as relative uniformity in density and absorption. Additionally, these methods assume that the depth in the X-ray sample used to develop the model is the same as for the X-ray catalog being crossmatched to. Quantities such as X-ray depth and absorption vary across the sky, making a reliable all-sky crossmatch difficult.

Compounding this fact, the upcoming Legacy Survey of Space and Time (LSST) by the Vera C. Rubin Observatory \citep{Ivezic:2019:LSST} will provide the deepest all-sky optical catalog to date. LSST will observe more than 18,000 sq. deg.  of the southern sky in six broadband filters ($u$, $g$, $r$, $i$, $z$, $y$) to single-visit depths of 24.5 mag and co-added depths of 27.5 mag. A crossmatch between LSST and the latest X-ray all-sky surveys, such as \eROSITA, has the potential to uncover vast numbers of rare objects. The high density of optical sources will make building a reliable set of unambiguous matches more difficult than ever. However, with roughly 3-4 days between visits, the cadence of LSST will enable an orthogonal approach towards evaluating potential optical counterparts, as each candidate will include not only color information, but time domain data as well. Many X-ray emitting objects are also transient or variable at optical wavelengths (see table \ref{tab:xray_optical_variability}) and we thus aim to leverage this fact to aid in the identification of optical counterparts.

To evaluate the difficulty of performing an X-ray/optical crossmatch at LSST depths, as well as the concept of using variability to aid in counterpart identification, we perform a crossmatch between a combined X-ray catalog of sources from several X-ray missions, and the first data preview from Rubin Observatory. The Rubin Observatory Data Preview \citep[DP1;][]{Rubin:DP1}, released June 30, 2025, is the first on sky ``science-grade'' dataset and the second of three data previews from LSST. It comprises about 2,000 exposures from seven fields captured with the LSST Commissioning Camera (LSSTComCam) in November and December 2024. DP1 includes raw and pipeline-processed visit images, visit-level source catalogs, co-added images with object catalogs, forced source catalogs, difference images and difference image analysis (DIA) catalogs, and other Science Support Products delivered via the Rubin Science Platform for early-science testing. This preview allows the community to validate the pipelines, explore real early data, and prepare for the full LSST survey.

In Section \ref{sec:methods}, we present the X-ray catalogs we selected and combined, as well as the contents of and data products we utilized in DP1. We also discuss challenges of performing the crossmatch to dense optical fields and how we utilize \textsc{lsdb} \citep[][]{Caplar:2025:LSDB} to perform the crossmatch. In Section \ref{sec:results}, we present the crossmatched catalog and evaluate the most likely optical counterparts. In Section \ref{sec:disc} compare to crossmatches to known sources in the DP1 fields, and examine where the most reliable matches fall in X-ray and optical feature space. In Section \ref{sec:concl}, we summarize our findings and look ahead to our expectations for LSST.

\begin{table*}[ht]
\centering
\begin{tabular}{ c c c c }  
    \hline \hline
\textbf{Object Type} & \textbf{X-ray Mechanism} & \textbf{Optical Signature} & \textbf{Timescale} \\
\hline
Flare Star & Coronal reconnection & Flares & Minutes-hours  \\
Active Binary & Coronal reconnection & Ellipsoidal variations, eclipses & Hours-days \\
\rev{YSOs} & \rev{Magnetic activity, accretion} & \rev{Flares, bursts, dips} & \rev{Hours-days} \\Cataclysmic Variable & White dwarf accretion & Outbursts, eclipses & Hours-days (outbursts) \\
\rev{Low Mass} X-ray Binary & NS/BH accretion & Ellipsoidal variations, outbursts & Hours-days (orbital), \\
 & & & Weeks-months (outbursts)  \\
\rev{High Mass X-ray Binary} & \rev{NS/BH wind accretion} & \rev{Pulsations, periodic brightening} & \rev{Hours-days (pulsations),}\\
 & & \rev{super-orbital modulation} & \rev{days-years (orbital)}  \\
 & & & \rev{Weeks-months (super-orbital)} \\
AGN / Quasar & Accr. Disk corona around SMBH & Aperiodic variability & Days-years  \\
TDE & Stellar disruption by BH & Power-law decay & Days-months \\
\rev{Galaxies}  & \revtwo{Bremsstrahlung, line emission,} & - &  - \\
 & \revtwo{and X-ray binary accretion} &  &   \\
\rev{Galaxy Clusters} & \rev{Thermal bremsstrahlung} & - & - \\
\hline
\end{tabular}
\caption{Summary of X-ray emitting objects with their optical variability features and timescales.}
\label{tab:xray_optical_variability}
\end{table*}

\section{Methods}
\label{sec:methods}

\begin{figure}[t]
    \centering
    \includegraphics[width=0.5\textwidth]{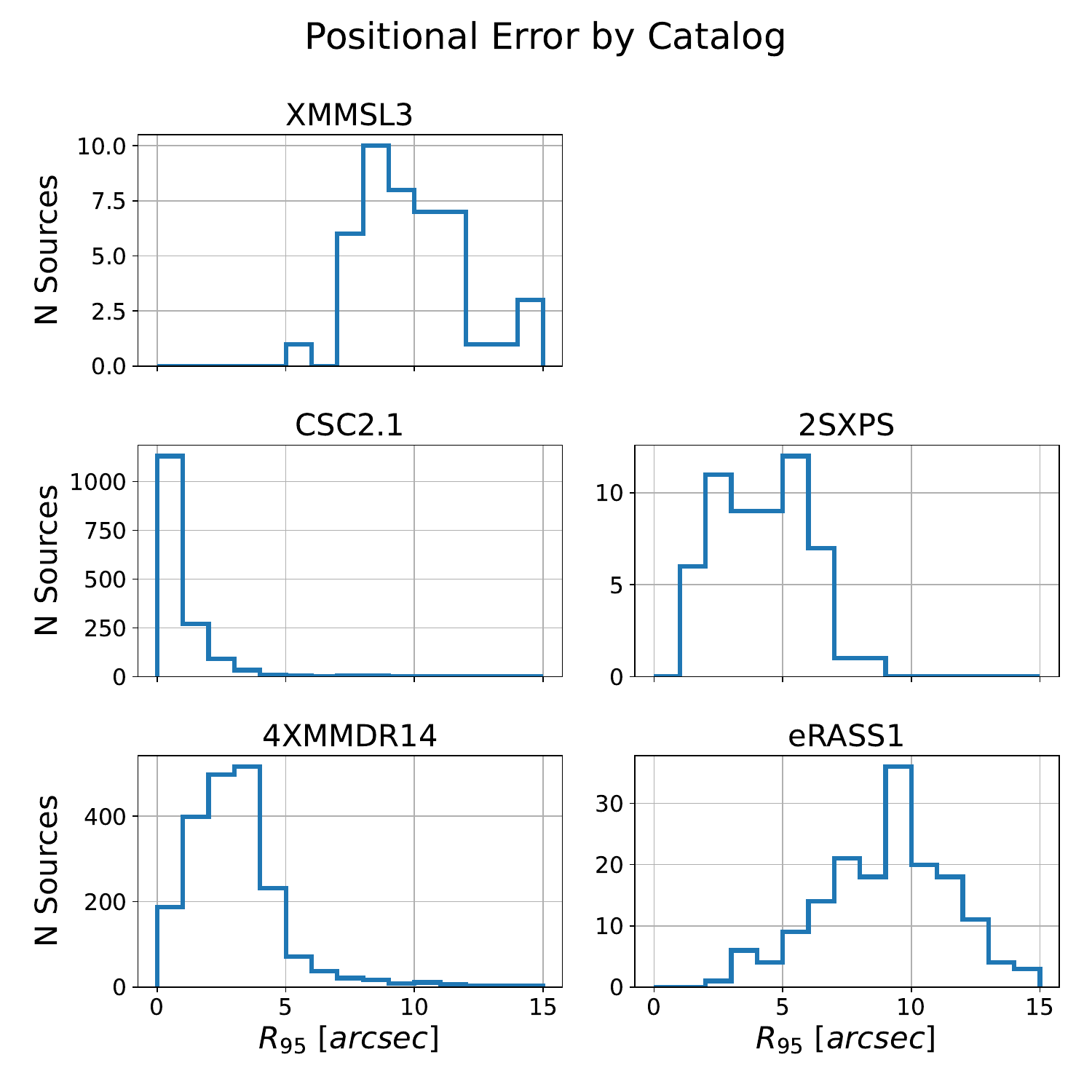}
    \caption{Histogram of $R_{95}$ positional errors of all X-ray sources in DP1 observed fields by catalog.}
    \label{fig:errors_hist}
\end{figure}

\begin{table*}
    \centering
    \begin{tabular}{ c c c c }  
        \hline \hline
        \textbf{Catalog} & \textbf{Column} & \textbf{Significance / Range} & \textbf{Notes} \\
        \hline
        \multirow{2}{*}{CSC2.1} 
            & err\_ellipse\_r0 & $1.960\sigma$ (95\%) & Major axis of the error ellipse \\
            & flux\_aper90\_avg\_b & 0.5–7.0 keV & Background-subtracted, aperture-corrected, 90\% ECF aperture \\
        \hline
        \multirow{2}{*}{4XMMDR14} 
            & SC\_POSERR & $0.933\sigma$ (63\%) & Weighted average of the total POSERR for source \\
            & SC\_EP\_8\_FLUX & 0.2–12.0 keV & Mean, error-weighted combined band flux \\
        \hline
        \multirow{2}{*}{XMMSL3} 
            & radec\_err & $1\sigma$ (68\%) & Statistical error on source position from detection software \\
            & flux\_b8 & 0.2–12.0 keV & Total energy band flux \\
        \hline
        \multirow{2}{*}{2SXPS} 
            & Err90 & $1.645\sigma$ (90\%) & 90\% confidence radial position error on the source position \\
            & PowFlux & 0.3–10.0 keV & From power-law indicated by WhichPow \\
        \hline
        \multirow{2}{*}{eRASS1} 
            & RADEC\_ERR & $1\sigma$ (68\%) & Combined positional error, raw output from PSF fitting \\
            & ML\_FLUX\_P[1..5] & 0.2–8.0 keV & Sum of bands 1–5 where DET\_LIKE\_P[band] $\geq 7$ \\
        \hline
    \end{tabular}  
    \caption{Positional uncertainty and flux column details for each X-ray catalog.}
    \label{table:catalogs}
\end{table*}

\subsection{X-ray Catalog} \label{sec:xray_cat}
We began by assembling a combined X-ray catalog from the most up-to-date releases for five X-ray catalogs: the \Chandra{} source catalog \citep[CSC2.1;][]{Evans:2024:CSC2.1}, the \XMM{} serendipitous source catalog \citep[4XXM-DR14;][]{Webb:2020:4XMMDR14}, the \XMM{} slew survey catalog \citep[XMMSL3;][]{Saxton:2008:XMMSL}, the Swift-XRT point source catalog \citep[2SXPS;][]{Evans:2020:2SXPS}, and the \eROSITA{} all-sky survey source catalog \citep[eRASS DE DR1 (hereafter just eRASS1);][]{Merloni:2024:eRASS1}. From each catalog, we selected the name, right ascension, declination, positional uncertainty, and broadband X-ray flux. 

\new{The column names for the broadband flux and positional uncertainty are listed in Table \ref{table:catalogs}. In each case we choose the column that accounts for both systematic uncertainty and positional uncertainty from source detections. For \Chandra{}, we take the major axis of the error ellipse as our uncertainty radius. We take this uncertainty and scale it to a $1\sigma$ error radius $R_{1\sigma}$. Then, assuming Gaussian errors, we multiply to get the 95\% error radius $R_{95} = 1.96R_{1\sigma}$.} 

\new{Since the energy range of the broadband flux is different for each instrument, we list the energies in Table \ref{table:catalogs}, as well as an additional description of the flux pulled from the documentation for each catalog. The eRASS1 broadband flux was calculated by adding bands 1 to 5 for each X-ray source, provided the source is detected with DET\_LIKE\_P[band] $\geq 7$, for a moderate strength detection in that band. The distribution of positional errors in each catalog are plotted in Figure \ref{fig:errors_hist}.}

\begin{figure}[ht]
    \centering
    \includegraphics[width=0.49\textwidth]{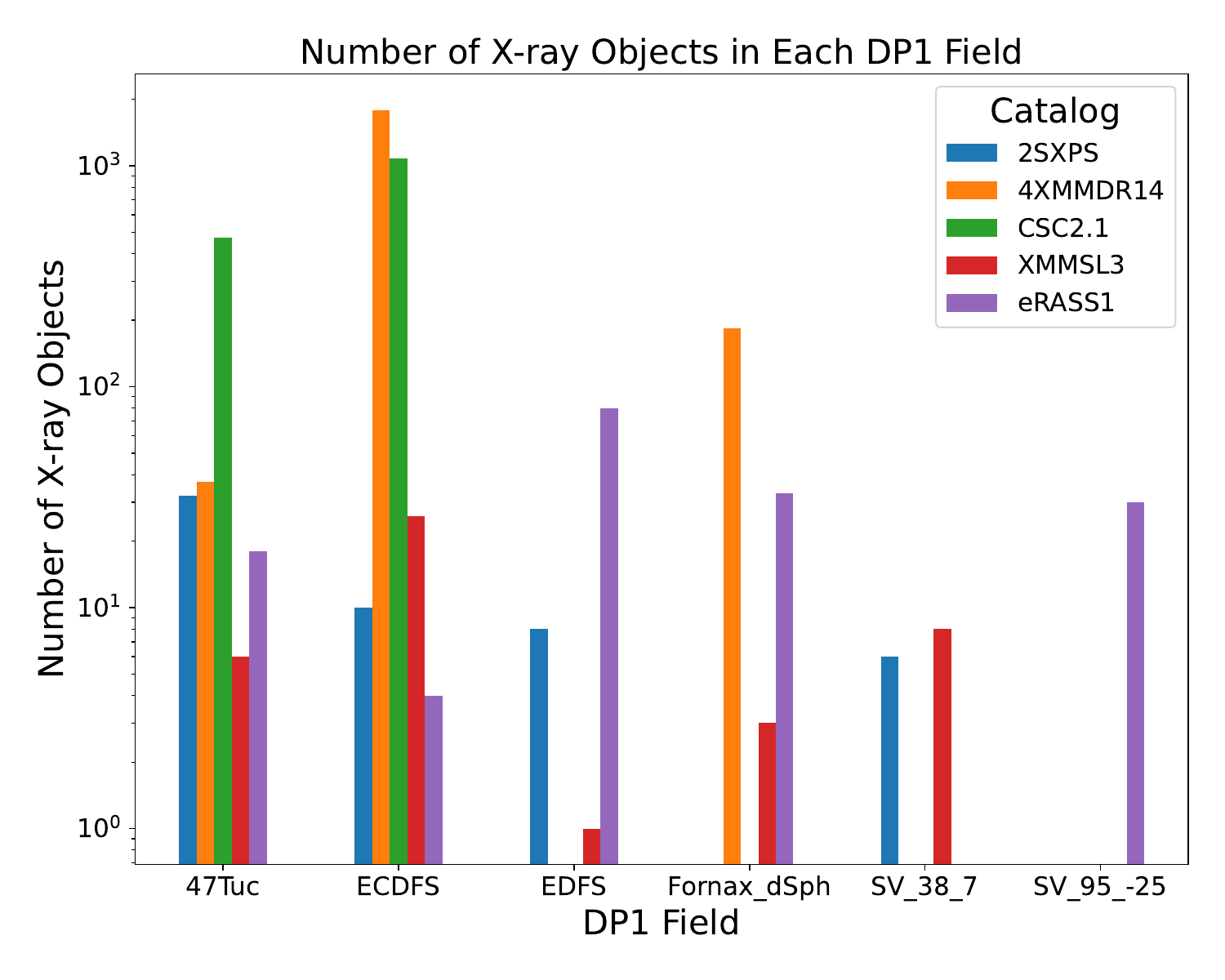}
    \caption{Number of X-ray sources by X-ray catalog in each DP1 field. The majority of X-ray sources are found in the Extended Chandra Deep Field - South (E-CDF-S), which is well observed by X-ray telescopes.}
    \label{fig:xray_dp1_field}
\end{figure}

We then performed a 1.5 degree conesearch around the center of each of the 6 DP1 fields we include in our crossmatch to isolate only the X-ray sources near these fields, yielding \new{10,307} sources. We then de-duplicated the catalogs by crossmatching between all pairs of X-ray catalogs to get all sources that fall within the 95\% error circle for another source. For 1:1 matches, we assume the two sources are the same and keep the source with the smaller positional uncertainty. For sources with multiple matches, we keep only the sources from the catalog with the smaller positional uncertainty. This leaves \new{9248} sources. We finally impose a 15$^{\arcsec}$ limit on the 95\% error radius, which eliminates outliers and leaves 9089 total sources. 

Finally, \rev{we} eliminate all sources that are never observed as part of DP1. \rev{The data processing pipelines divide the celestial sphere into tracts, with each tract covering approximately $2.8\,\mathrm{deg}^2$. Each tract is then further divided into 100 equally sized patches in a 10 by 10 grid \citep{Rubin:DP1}.} We find the tract and patch of each position using the DP1 \texttt{skymap}, leaving a total of \new{4481} final sources in our X-ray catalog. The number of X-ray sources in each DP1 field, broken out by X-ray survey, is given in Figure \ref{fig:xray_dp1_field}. 

\subsection{Optical Catalog} \label{opt_cat}
The LSST Data Preview 1 (DP1) contains the first data collected by the Vera C. Rubin Observatory and offers the first look at the characteristics and capabilities of the data that will eventually be in LSST. The dataset consists of images taken by LSSTComCam between November 9 and December 12, 2024 and covers an area of approximately 15 sq. deg. We selected six of the fields observed by DP1: 47 Tucanae (47 Tuc), the Extended Chandra Deep Field South (E-CDF-S), the Euclid Deep Field (EDF-S), the Fornax Dwarf Spheroidal Galaxy (Fornax dSph), and two science validation fields, one at low galactic latitude (SV\_95\_-25) and one at low ecliptic latitude (SV\_37\_7) \citep{Rubin:DP1}, excluding the Seagull Nebula field which contains extensive nebulosity. These fields were observed repeatedly in the $ugrizy$ bands over several nights. Observations were made in observing epochs on a given target field, with 5--20 visits in each of three loaded filters at a time. Each visit consists of one 30 second exposure covering approximately half a square degree. 

The spatial distribution and coverage of the DP1 fields, and the temporal distribution of observations by field and filter can be found in the DP1 science paper \citep[][figures 4-6]{Rubin:DP1}. The \rev{fields were not observed equally and in particular, the E-CDF-S field approaches 10 year survey depths. The number of visits per filter} is given in Table \ref{table:visits}. The cumulative imaging depth at S/N = 5 expressed as a limiting magnitude in each field exceeds 26th in the most observed regions in each field and is visualized in figure 3 of the \textit{Rubin Observatory Plans for an Early Science Program} document \citep{RTN-011}.

For DP1 data, we adopt the Rubin convention of referring to a single-epoch detections within science images as a \Source{} and the astrophysical object associated with a given detection as an \Object. To draw this distinction, we refer to Rubin data products such as \Source{}s and \Object{}s in teletype for the remainder of the paper. In other words, every \Object{} will usually have multiple associated \Source{}s, with each \Source{} corresponding to an observation in one single-epoch science image, also known as a \visitIm. DP1 consists of five types of data products, which are described in detail in \cite{Rubin:DP1}: images, catalogs, maps, metadata, and ancillary data products. We utilize the \Object{} catalog \citep{object}, which contains 2,299,726 \Object{}s---detections on \deepCoadd{} images \citep{deepcoadd} made by stacking all available \visitIm{}s \cite{visit_image}. To produce lightcurves and calculate time-series features, we require per-visit photometry. We use the \ForcedSource{} catalog \citep{ForcedSourceOnObject}, which contains forced photometry at the position of detected \Object{}s in each \visitIm. 

\new{The Rubin difference image analysis (DIA) pipeline also detects sources in \diffIm{}s \citep{difference_image}, known as \DiaSource{}s \citep{dia_source}. By associating \DiaSource{}s at the same position, the DIA pipeline also builds \DiaObject{} catalogs. We also crossmatch to this catalog, which contains 1,089,818 \DiaObject{}s \citep{dia_object}. The DIA pipeline bypasses the deblending step that fails in the densest crowded fields \citep{Choi:2025:47TucDP1}, such as inner regions of the 47 Tuc and Fornax dSph fields. Other studies have shown that matching to the \DiaObject{} table can recover photometry from some of the denser fields \citep{Wainer:2025:47Tuc}. More discussion of the \DiaObject{} catalog match is found in \ref{sec:dia}.}

\revthree{DP1 catalogs are a proprietary data product with access specified by the LSST data rights policy \citep{RDO-013}. Rubin data rights holders can run our code, available at \cite{wang_2026_18665437}, on the Rubin Science Platform (RSP) to replicate our analysis.}

\subsection{Crossmatch to \Object{} table}
\textsc{lsdb} was developed by the LSST Interdisciplinary Network for Collaboration and Computing (LINCC) Frameworks project to enable scalable analysis of large catalogs \citep{Caplar:2025:LSDB}. We use the \textsc{lsdb} \new{Hierarchical Adaptive Tiling Scheme (HATS) formatted DP1 \Object{} catalog \citep{Malanchev:2025:LSDB_DP1} and the \textsc{lsdb}} crossmatch function to crossmatch the combined X-ray catalog to all \Object{}s detected in the six DP1 fields. We set the match radius to 15$^{\prime\prime}$ and return up to the 10 closest matches for each X-ray source. We then eliminate all matches where the candidate optical counterpart does not fall within the 95\% error radius of the X-ray source. We are left with \new{4513} candidate \Object{} counterparts for \new{2314} unique X-ray sources. 

\new{Since we had narrowed down our catalog of X-ray sources by keeping all sources in tracts and patchs observed by LSSTComCam, some of these sources are just past the outskirts of the DP1 fields and are not in the observed footprint.} To further eliminate these sources, we find the convex hull\rev{, the smallest convex polygon that encloses} all matches in \rev{each} field. We then assume that sources interior to the convex hull were observed as a part of DP1 and are therefore optical nondetections. This eliminates \new{651} of the X-ray sources and leaves us with a final total of \new{3830} X-ray sources in the footprint of DP1. 

\new{Lastly, we repeat the crossmatching, adding 30$\arcsec$ to the declination of each X-ray source to obtain a random set of matched \Object{}s, so we can compare our matches to a control sample of field objects.} 

\begingroup 
    \setlength{\tabcolsep}{7pt} 
    \renewcommand{\arraystretch}{1.2} 
    \setlength\extrarowheight{1.5pt}
    \begin{table}
        \centering
        \begin{tabular}{ c c c c c c c }  
            \hline \hline
            {\boldmath Target} & {\boldmath u} & {\boldmath g} & {\boldmath r} & {\boldmath i} & {\boldmath z} & {\boldmath y} \\
            \hline 
            47 Tuc & 6 & 10 & 33 & 19 & 0 & 5 \\
            Rubin SV 38 7 & 0 & 44 & 55 & 57 & 27 & 0 \\
            Fornax dSph & 0 & 5 & 26 & 13 & 0 & 0 \\
            E-CDF-S & 53 & 230 & 257 & 177 & 177 & 30 \\
            EDF-S & 20 & 61 & 90 & 42 & 42 & 20 \\
            Rubin SV 95 -25 & 33 & 86 & 97 & 29 & 60 & 11 \\
            \hline  
        \end{tabular}  
        \caption{Number of visits by field for DP1.}
        \label{table:visits}
    \end{table}
\endgroup

\begin{table*}[ht]
\centering
\begin{tabular}{llccccccc}
\toprule
\multirow{2}{*}{Feature} & \multirow{2}{*}{Subset} & \multicolumn{6}{c}{Field} & \multirow{2}{*}{Total} \\
\cmidrule(lr){3-8}
& & 47 Tuc & E-CDF-S & EDF-S & Fornax\_dSph & SV\_38\_7 & SV\_95\_-25 & \\
\midrule
\rev{X-ray sources} & \rev{…} & \rev{567} & \rev{2910} & \rev{89} & \rev{220} & \rev{14} & \rev{30} & \rev{3830} \\
\hline \\
X-ray sources with \\ \Object{} match & 2SXPS & 0 & 8 & 8 & 0 & 6 & 0 & 22 \\
 & 4XMMDR14 & 4 & 1279 & 0 & 2 & 0 & 0 & 1285 \\
 & CSC2.1 & 3 & 821 & 0 & 0 & 0 & 0 & 824 \\
 & XMMSL3 & 2 & 24 & 1 & 3 & 8 & 0 & 38 \\
 & eRASS1 & 10 & 4 & 76 & 27 & 0 & 28 & 145 \\
 & Total & 19 & 2136 & 85 & 32 & 14 & 28 & 2314 \\
\\
X-ray sources with \\ \DiaObject{} match  & ... & 91 & 810 & 64 & 71 & 10 & 21 & 1067 \\
\\
Any \Object{} or \\ \DiaObject{} match  &  ... & 101 & 2241 & 89 & 91 & 14 & 30 & 2566 \\
\\
No \Object{} or \\ \DiaObject{} match  & ... & 466 & 669 & 0 & 129 & 0 & 0 & 1264 \\
\hline \\
$N_\Object$ in $R_{95}$ & ...  & 75 & 3560 & 429 & 195 & 57 & 197 & 4513 \\
\\
$N_\DiaObject$ in $R_{95}$ & ... & 249 & 1587 & 176 & 106 & 31 & 127 & 2276 \\
\hline \\
Crossmatch Reliability & $>90\%$ & 8 & 1295 & 21 & 10 & 3 & 7 & 1344 \\
\hline \\
\texttt{refExtendndess}  & point-like (0) & 7 & 379 & 23 & 14 & 1 & 8 & 432 \\
 & extended (1) & 7 & 1617 & 53 & 16 & 10 & 11 & 1714 \\
 & unknown (NA) & 5 & 140 & 9 & 2 & 3 & 9 & 168 \\
\hline \\
External xmatches & Gaia & 10 & 235 & 45 & 19 & 8 & 28 & 345 \\
 & Milliquas & 2 & 434 & 3 & 0 & 5 & 0 & 444 \\
 & SIMBAD & 1 & 1036 & 2 & 1 & 2 & 4 & 1046 \\
\bottomrule
\label{tab:breakdown}
\end{tabular}
\caption{Breakdown of matches by field. The first section details the number of X-ray sources with DP1 \Object{} matches \rev{(detections in \deepCoadd{}s)} for each X-ray catalog, and the X-ray sources with matches to \DiaObject{}s \rev{(detections in \diffIm{}s)}. Since some sources are only detected in either the \deepCoadd{}s or \diffIm{}s, as seen in Figure \ref{fig:spatial_plot}, we also present numbers of X-ray sources with any \Object{} or \DiaObject{} match, as well as those with no match. The second section gives the total numbers of \Object{}s and \DiaObject{}s found in the 95\% error radius of any X-ray source---some sources having multiple matches. We also show the number of high reliability matches ($>90$\%), the number of point-like vs. extended \Object{}s, and the number of \Object{}s with matches to external catalogs}
\end{table*}

\subsection{Evaluation of crossmatch reliability}
We calculate the reliability\footnote{The crossmatch reliability should not be confused with the Rubin-reported reliability score for \DiaSource{}s, which is a machine-learned probability that a given \DiaSource{} is astrophysical in origin.}, which we define as the probability of finding no chance object closer to the X-ray source than the counterpart \Object{} \citep{Sutherland:1992:LR}. The expected number of background sources within the separation between the source and counterpart is:
\begin{equation}
Y = \pi d^2 \cdot N \text{,}
\label{eq:expected_background}
\end{equation}
where item \( Y \) is the expected number of background sources, \( d \) is the separation between the source and counterpart, and \( N \) is the local background source density \citep{Sutherland:1992:LR}. We calculate the background by counting the number of \texttt{Objects} in an annulus with an inner radius of the 4 $\sigma$ X-ray error circle, $R_{4\sigma}$, and outer radius of $R_{8\sigma}$ around each counterpart. \new{We obtained these radii by scaling up the $R_{1\sigma}$ calculated in Section \ref{sec:xray_cat}.} We set the inner radius to $R_{4\sigma}$ in order not to bias the number density by including the real counterpart. If there are no \Object{}s in the annulus, we increase the outer radius by $10\arcsec$ until we have at least one. Assuming Poisson statistics, the reliability of the match is given by

\begin{equation}
\text{Reliability} = \text{Pr}(0, Y) = e^{-Y}.
\label{eq:rel_poisson}
\end{equation}

\subsection{Quantifying variability}

We also calculate the Stetson \( J \) index to quantify the variability of each potential counterpart \citep{Stetson:1996:StetsonJ}.
Let the standardized residual for observation \( i \) be defined as

\begin{equation}
\delta_i = \sqrt{\frac{n}{n - 1}} \cdot \frac{\nu_i - \bar{\nu}}{\sigma_i},
\label{eq:standardized_residual}
\end{equation}
where \( \nu_i \) is the observed value (e.g., flux or magnitude), \( \bar{\nu} \) is the weighted mean value, \( \sigma_i \) is the uncertainty in \( \nu_i \), and \( n \) is the total number of observations in the band. Since each LSSTComCam epoch consists of a single observation, we define the product $P_i$

\begin{equation}
P_i = \delta_{i}^2 - 1.
\label{eq:paired_product}
\end{equation}
The Stetson \( J \) index is then defined as

\begin{equation}
J = \frac{1}{N} \sum_{i=1}^{N} \text{sgn}(P_i) \cdot \sqrt{|P_i|}.
\label{eq:stetson_j_basic}
\end{equation}

We calculate the Stetson \( J \) index in each band using the \texttt{psfFlux} and \texttt{psfFluxErr} columns from the forced source table for each \texttt{Objects}. 

\subsection{Value added crossmatches to SIMBAD, Milliquas, Gaia}
After we obtained the crossmatch between DP1 \Object{}s and our X-ray catalog, we also matched these \Object{}s to several other catalogs to provide additional context to our matches and to build up a value-added catalog. \new{We used \textsc{lsdb} to find all Gaia DR3 sources \citep{Gaia:2023:DR3}, within 0.5$\arcsec$ of the LSST DP1 objects.} Since Gaia provides parallaxes and distances, we can use the data to distinguish between Galactic and extragalactic objects for some of the brighter \Object{}s. \new{We also find all DP1 sources within 2$\arcsec$ of any Gaia source brighter than 15th magnitude---these \Object{}s are flagged because they may either be saturated, or have inconsistent photometry due to the nearby presence of a bright star.}

We also crossmatched the positions of the optical LSST DP1 objects to the Million Quasar (Milliquas) catalog \citep{Flesch:2023:Milliquas}, using radius of 1$^{\prime\prime}$ since most of the objects in the catalog are sourced from optical surveys with high precision, such as Gaia eDR3, DESI, and SDSS-DR18. We expect many of our matches in the E-CDF-S and EDF-S fields, which are away from the galactic plane, to be extragalactic, and this crossmatch will enable us to compare the properties of known active galaxies to our broader set of matches.

Lastly, we used astroquery \citep{Ginsburg:2019:Astroquery} to query any existing object type (otype) from SIMBAD \citep{Wenger:2000:SIMBAD} within 1$^{\prime\prime}$ of the position of the \Object{} location, and any associated photometry in the $ugri$ \new{and $V$ bands. We then drop duplicate matches such that all remaining matches are between one SIMBAD source and one DP1 \Object{}.}

\section{Results} \label{sec:results}

\subsection{Crossmatch results}
\begin{figure*}[ht]
    \centering
    \includegraphics[width=0.97\textwidth]{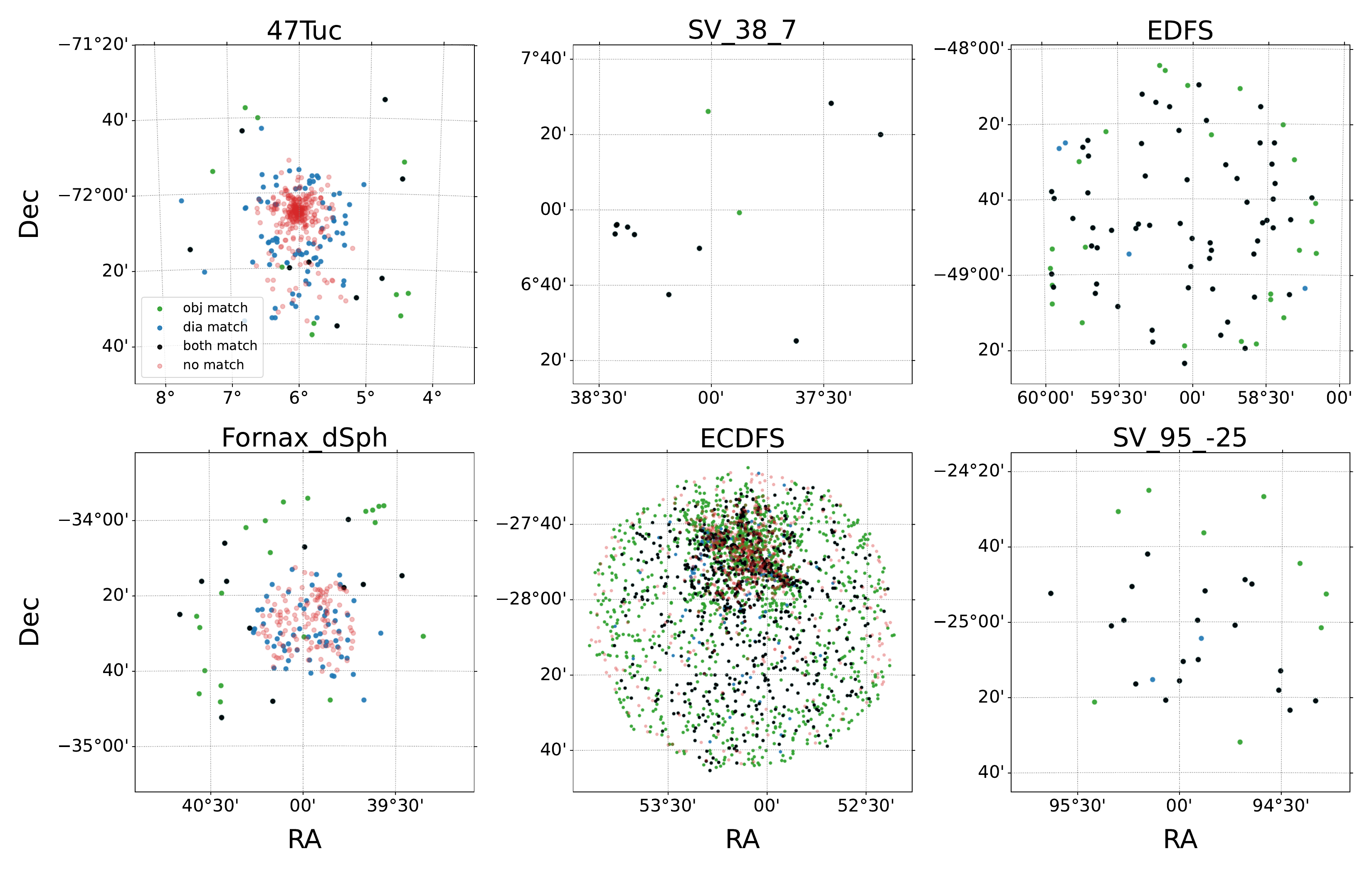}
    \caption{Spatial distribution of all X-ray sources in the combined X-ray catalog. Matches to \texttt{Objects} are colored green. We also crossmatched \texttt{diaObjects}, which are detections based on \diffIm{}s. X-ray sources only detected in the \DiaObject catalog are plotted in blue and sources found in both catalogs are plotted in black. X-ray sources with no optical match in DP1 are plotted in red. Due to the crowdedness of the 47 Tuc and Fornax dSph fields, we have better success matching to the \DiaObject{} table (see Section \ref{sec:dia}). Many of the unmatched sources are also in the cores of these two fields for the same reason.}
    \label{fig:spatial_plot}
\end{figure*}

\new{A total of 4513 candidate \Object{}s are found within the 95\% error radius for 2314 unique X-ray sources, as some X-ray sources have multiple DP1 \Object{}s within the $R_{95}$}. Figure \ref{fig:spatial_plot} shows the locations of all objects in the original combined X-ray catalog from Section \ref{sec:xray_cat} that were observed in the DP1 fields, as well as whether a counterpart was identified in DP1. 1600 X-ray sources did not have a counterpart in the DP1 \Object{}s table. Many of the unmatched X-ray sources lie in the cores of 47 Tuc (466/567) or the Fornax dSph galaxy (129/220), where the LSST pipeline performance is reduced due to the high stellar density \new{and there are no \Object{}s detected}. Additionally, 669 out of 2910 sources in the E-CDF-S field have no match in our crossmatched dataset. We find matches for all X-ray sources in the EDF-S field, the low galactic latitude science validation field (SV\_95\_-25), and the low ecliptic latitude field.

\begin{figure*}[ht]
    \centering
    \includegraphics[width=0.98\textwidth]{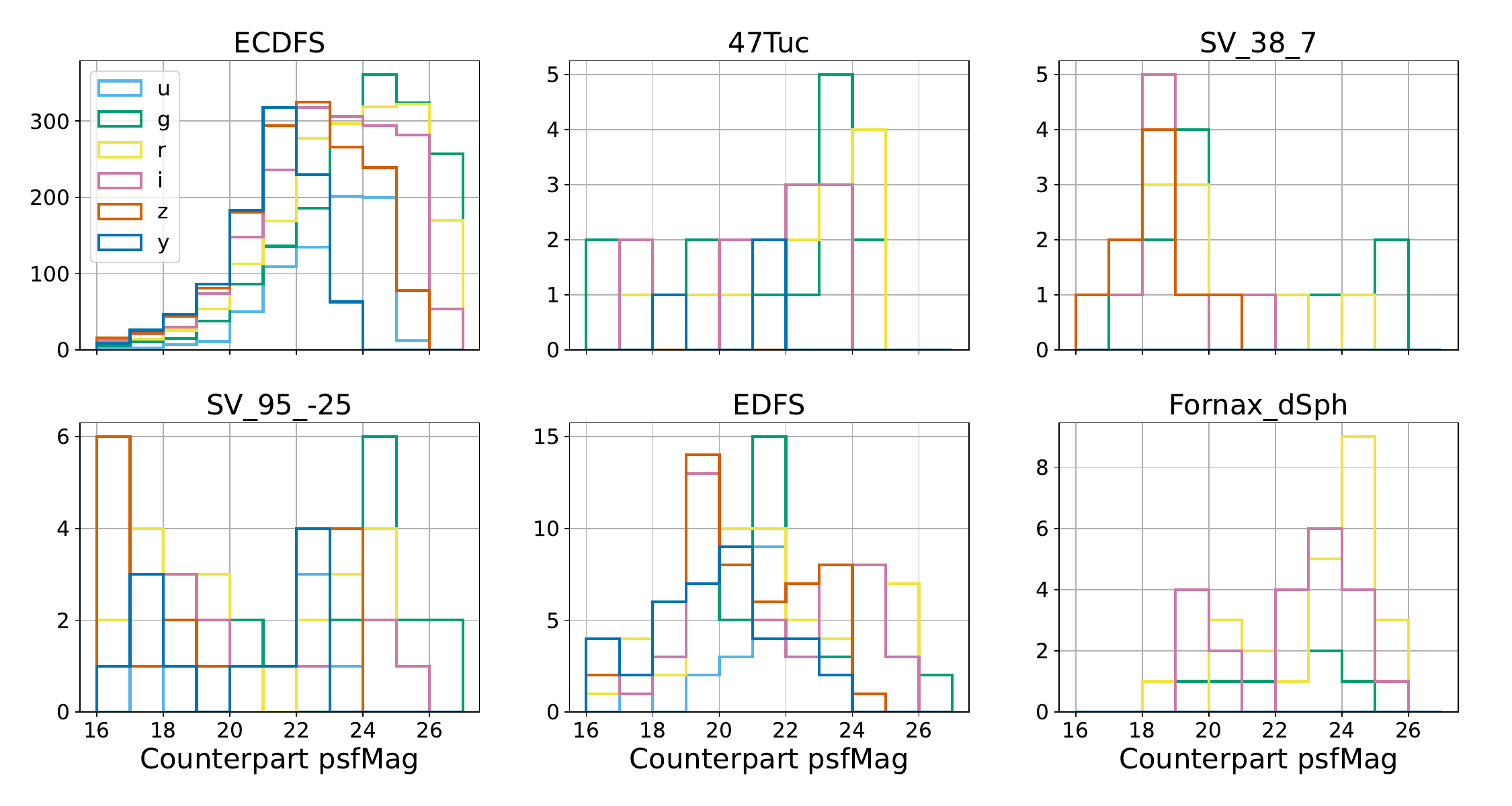}
    \caption{The magnitude distribution of the counts of candidate counterparts in each of the 6 fields, by band. Many of the crossmatched counterparts are fainter than the Gaia, PanSTARRS and SDSS limiting magnitudes. Only measurements where the SNR $\geq 5$, where SNR = \texttt{psfFlux}/\texttt{psfFluxErr} are in the histograms.}
    \label{fig:matches_mag}
\end{figure*}

The magnitude distribution of the counterparts in each field can be seen in Figure \ref{fig:matches_mag}. Taking fiducial limiting magnitudes of 21, 23, and 24 for Gaia \citep{Gaia:2023:DR3}, SDSS \citep{York:2000:SDSS}, and PanSTARRS \citep{Kaier:2004:PanSTARRS}, we find that 78\%, 45\%, and 27\% of DP1 objects were fainter than these surveys, respectively, in all bands. This indicates that during the full survey, we should expect to vastly increase the number of X-ray counterparts. 

The total numbers of matches broken out by field and instrument are given in Table \ref{tab:breakdown}. We also present the number of matches to external catalogs: Milliquas, Gaia, and SIMBAD, by field. We find a total of 444 matches between our DP1 \Object{}s and the Milliquas catalog, 345 between our DP1 \Object{}s and Gaia DR3, and 1046 unique matches between our DP1 \Object{}s and SIMBAD.

\subsection{Reliability of crossmatch}

Next, we look at the reliability of crossmatch for the most likely counterpart for each unique X-ray source. The reliability is influenced both by the radius of separation between the X-ray source and DP1 match and the local density. As expected, the fields with the most constrained X-ray observations - those made by \Chandra{} and \XMM{} - have the highest proportion of high reliability matches. 

\begin{figure}[ht]
    \centering
    \includegraphics[width=0.47\textwidth]{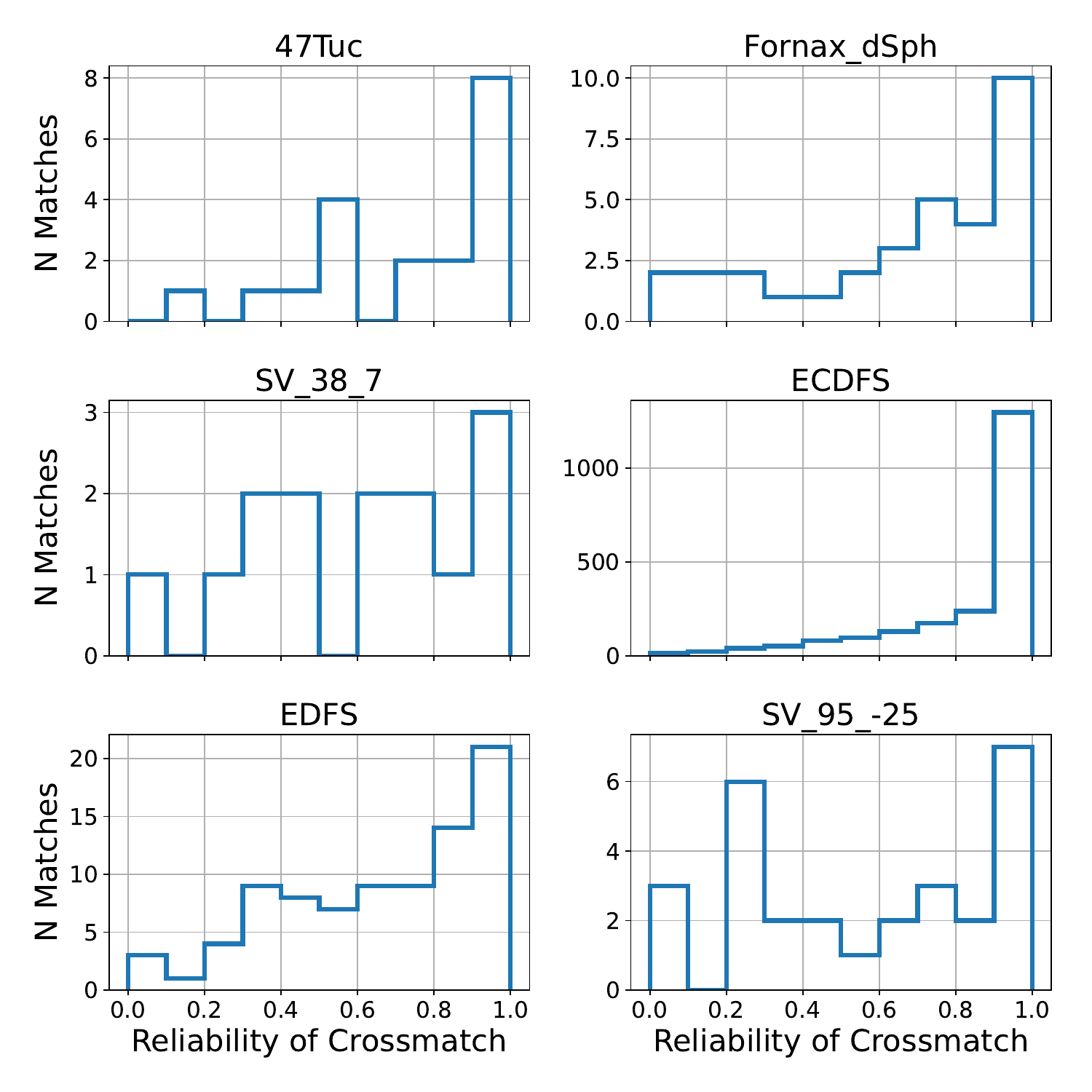}
    \caption{The reliability of crossmatch of the closest counterpart to each X-ray source, broken out by field. The reliability is the probability the counterpart is not a random association.}
    \label{fig:reliability_field}
\end{figure}

Figure \ref{fig:reliability_field} plots the distribution of reliability in each field. We define a match where the probability of a chance association is less than 10\%, i.e. the reliability $\geq 90\%$ to be a high-reliability match and we create a subset of 1344 matches meeting this criteria. The E-CDF-S field has the highest proportion of reliable matches, with a total of 1295, or 60.6\% of all matches in that field. On the other hand, the proportion of reliable matches is much lower in the EDF-S and the two science validation fields at between 21\% and 25\%. 

The calculation of the reliability depends on an accurate measurement of the background source number density. This in turn assumes that the \Object{} catalog has high completeness---otherwise the density will be underestimated and the reliability will be falsely high. Because we know that the two crowded fields 47 Tuc and Fornax dSph may not meet this requirement, we should treat the reliabilities calculated in these fields with caution.

\begin{figure}
    \centering
    \includegraphics[width=0.47\textwidth]{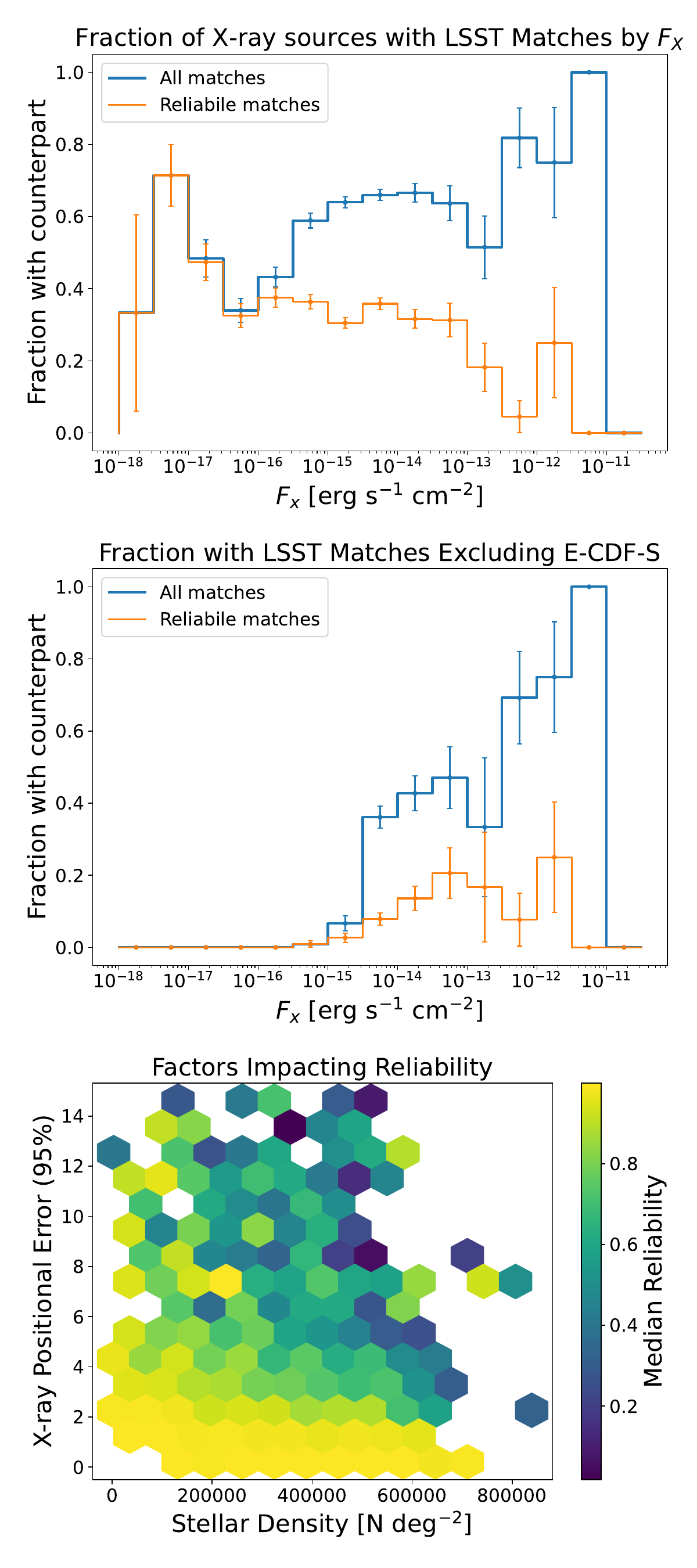}
    \caption{\rev{Top: The fraction of sources in each X-ray Flux bin with a match and with reliable match in the DP1 \Object{} catalog. The error bars show the standard error. The fainter X-ray sources are dominated by sources from the CSC2.1, which are generally well localized, leading to an increase in fraction of sources with reliable matches. Middle: Excluding the E-CDF-S field, which is both the best observed by X-ray observatories and by DP1, we see that the fraction falls off strongly with \rev{absorbed} X-ray flux. This is more indicative of what we expect a crossmatch to early Rubin (e.g. DR1) data may look like. Bottom: The main factors impacting our reliablity score are the X-ray positional error and the density of \Object{}s around the X-ray source.}}
    \label{fig:reliability_breakdown}
\end{figure}

\rev{We also find that due to the definition of our reliability criteria, the X-ray positional error and the local density of \Object{}s has the greatest impact on reliability, with well localized sources in less crowded areas having the highest reliability scores. Figure \ref{fig:reliability_breakdown} plots the fraction of X-ray sources with any counterpart in DP1, as well as the fraction of sources with a reliable counterpart. We see that the fraction of sources with a match and a reliable match tend to fall off with X-ray flux in non E-CDF-S fields. In these fields, most of the X-ray sources are from eROSITA and the XMM slew surveys, which are less sensitive and have comparatively poorer localizations when compared to sources in CSC2.1 and XMMDR14 catalogs, and there are very few sources fainter than $10^{-15}\,{\rm erg} {\rm s}^{-1} {\rm cm}^{-2}$. The uppermost subplot is dominated by matches in the E-CDF-S field, where most matches are to sources in the CSC2.1 and XMMDR14 catalogs, which are fairly well localized. This leads to an increase in sources with reliable matches at lower fluxes, as these sources tend also to be observed out of these catalogs.}

\section{Discussion}\label{sec:disc}

\subsection{Comparison to E-CDF-S crossmatch in \cite{Luo:2017:ECDFS}}

Out of 1055 total objects in the E-CDF-S X-ray catalog in \cite{Luo:2017:ECDFS}, we find 777 counterparts in the \Object{} catalog within the $2\sigma$ error radius and 900 counterparts in the $3\sigma$ error radius. Of the remaining 155 \Chandra{} sources, 79 are only detected in bands with wavelength longer than LSST y-band, or have no optical or IR counterpart in the original study either. The final 75 sources with Hubble Space Telescope (HST) counterparts in either R or z band only contains 21 sources brighter than 24th magnitude. We also have one match (\texttt{LSST-DP1-O-611255759837095315}) to one of the 16 X-ray sources with no match in the original paper (No 896). We find that $log(F_{X}/F_{r\_psfFlux}) = -0.84 > -1$, and the DP1 \texttt{refExtendedness} parameter for this \Object{} is 1, i.e. extended, which satisfies two of the criteria in the paper for an active galactic nucleus (AGN) classification. 

\subsection{Variability}
\begin{figure}
    \centering
    \includegraphics[width=0.47\textwidth]{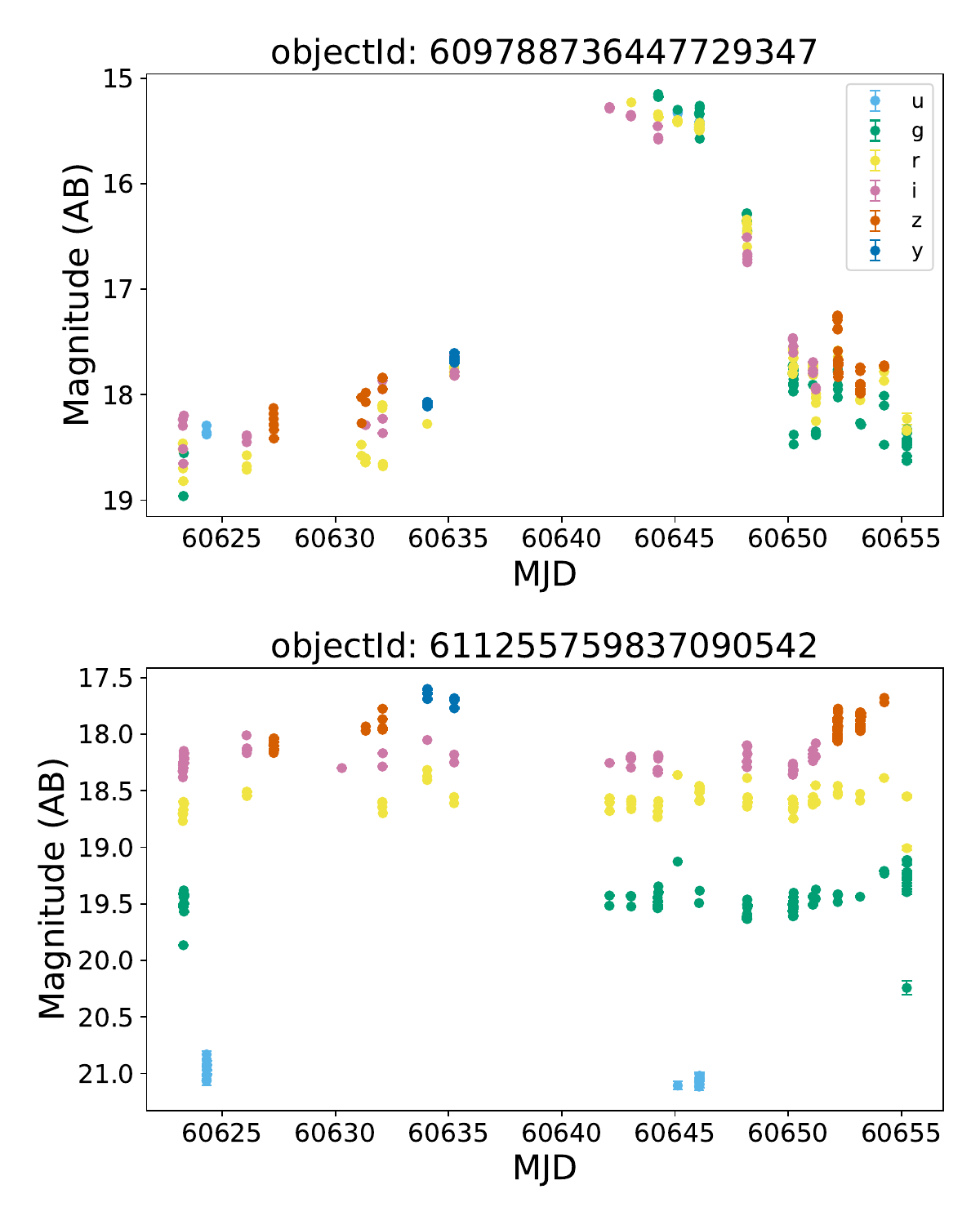}
    \caption{Example lightcurves of our counterparts. \rev{The top plot shows the SU UMa type cataclysmic variable CRTS J033349.8-282244, which was observed in outburst. This \Object{} had a $r$-band Stetson \( J \) of 450, highest out of all of our high-reliability matches. The bottom plot shows the AGN 3D-HST 38237, an AGN \revtwo{with a} $r$-band Stetson \( J \) of 24.}}
    \label{fig:cv_lightcurve}
\end{figure}

We also examine whether the Stetson \( J \) index has any relation to the reliability of the closest \Object{} counterpart, to test our hypothesis that the optical counterparts to X-ray sources will tend to be more variable. As an example, we plot the lightcurve of high-reliability match with the largest $r$-band Stetson \( J \) in Figure \ref{fig:cv_lightcurve}. This turns out to be the SU UMa type cataclysmic variable CRTS J033349.8-282244 \citep{Drake:2014:CRTS_CVs, Kato:2017:SU_UMa_CVs}, which completed a full outburst cycle during commissioning observations, and was also found by \cite{Malanchev:2025:LSDB_DP1} in their variability search. \rev{We also plot the lightcurve of one AGN, 3D-HST 38237 in the Milliquas catalog, with a fairly large Stetson \( J \). We note that for both objects, the intra-day variation in each band much greater than the estimated error, hinting at possible systematics in the PSF photometry of these bright objects.}

\begin{figure*}
    \centering
    \includegraphics[width=0.97\textwidth]{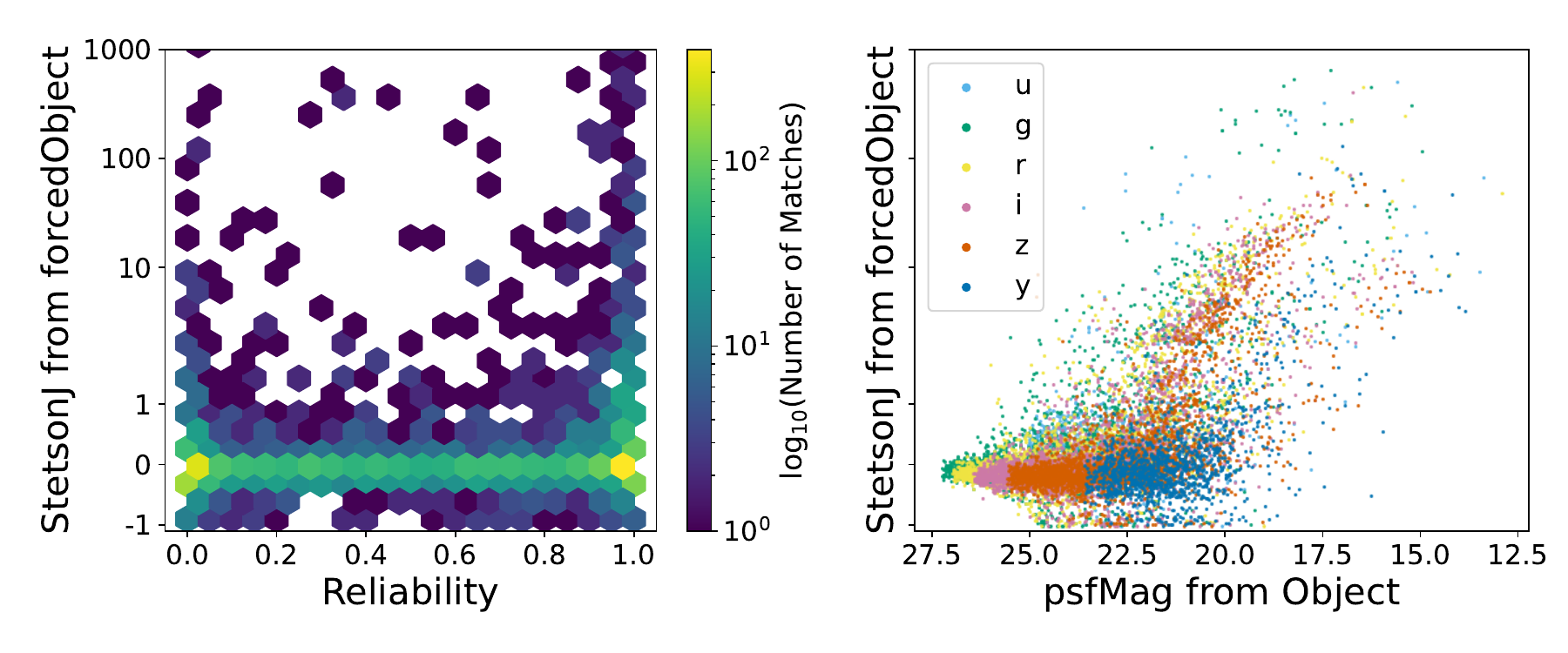}
    \caption{\rev{There may be a slight association} between the reliability of a crossmatch and the Stetson \( J \) index of the object, {as seen in the histogram on the left}. The Stetson \( J \) appears \rev{to increase with a subset of brighter \Object{}s, shown by the spur in the right plot. A possible explanation for the spur is a systematic underestimate of the photometric error.} We use a symmetric logarithmic (symlog) scale for the y-axis.}
    \label{fig:reliability_stetsonJ}
\end{figure*}

Plotting the reliability of crossmatch against the Stetson \( J \) index in Figure \ref{fig:reliability_stetsonJ}, we find that there appears to be no \rev{strong} association between these two metrics \rev{in the left subplot}. Instead, our Stetson \( J \) appears correlated with the magnitude of the \Object{}. \rev{Additionally, almost all objects brighter than 20th magnitude have Stetson \( J \) $>$ 1. The intra-day variation visible in each band in the lightcurves in figure \ref{fig:cv_lightcurve} suggest that systematics may play a role in the larger \( J \)s at brighter magnitudes. We expect these sources to be fairly stable in brightness over the course of each set of same-band observations, which are observed in blocks of 5-20 visits, lasting a total of less than an hour each.} 

We perform a 2-sample KS-test with the null hypothesis that the distribution of Stetson \( J \) for field \rev{\Object{}s} is the same or greater than as the distribution of Stetson \( J \) for our matches. Due to the magnitude dependence of the Stetson \( J \), we perform this test on samples for matches and field objects of magnitude 17 to 25 in bins of width 2 magnitudes. We also split the samples into $ugri$ bands, for a total of 12 tests. We set our overall threshold $\alpha = 0.05$ for rejecting the null hypothesis and apply the Bonferroni Correction to obtain a per-test threshold of $\alpha/m=0.0042$, where $m$ is the number of hypotheses. We find that since we reject the null hypothesis for only two of these subsets at the $p<0.0042$ level, so we have at best weak evidence that the matches are more variable than the field \Object{}s.

\rev{We repeat the procedure above, comparing the distribution of Stetson \( J \) index for reliable matches against the distribution for matches that do not meet our reliability criteria. Here, we find that we reject the null hypothesis for only one of these subsets at the $p<0.0042$ level, so we again have minimal evidence that the reliable matches are more variable than the unreliable matches.}

The dense cadence of the commissioning observations means the Stetson \( J \)s are most sensitive to variability on timescales of seconds to days, since the majority of the DP1 observations will be on these timescales. This differs from much of the expected longer variability timescales of most X-ray counterparts, particularly AGN (see table \ref{tab:xray_optical_variability}), which we expect to make up the vast majority of our counterparts. 

\rev{The optical variability of an AGN is commonly quantified with the (noise-subtracted) structure function (SF), which describes how the typical magnitude change grows with the time lag $\Delta t$ between two observations. For a light curve in magnitudes $m(t)$ with measurement uncertainties $\sigma_\textrm{noise}$, we take the definition of the structure function by \cite{diClemente:1996:AGN_SF}:}
\begin{equation}
\rev{\mathrm{SF}(\Delta t)\equiv \sqrt{\frac{\pi}{2} \left\langle \big(m(t+\Delta t)-m(t)\big) \right\rangle^2 \;-\; \left\langle \sigma_{\textrm{noise}} \right\rangle^2 },}
\end{equation}
\rev{where $\langle\cdot\rangle$ denotes an average over all pairs of epochs separated by $\Delta t$. The factor $\frac{\pi}{2}$ is included assuming that both the intrinsic variability and the observational noise follow Gaussian distributions. Intuitively, $\mathrm{SF}(\Delta t)$ is the intrinsic root-mean-square magnitude difference on timescale $\Delta t$, after accounting for photometric noise.}

\rev{\cite{DeCicco:2022:AGN_SF_VLT} used 3.3 years of VLT Survey Telescope (VST) COSMOS data to quantify AGN optical variability at LSST depths. The maximum baseline for \textit{Object}s in the DP1 dataset with a reliable counterpart to a known AGN is 32 days. Using the structure functions obtained from their data, we would expect a difference of around 0.1 magnitudes for a 32 day time difference.}

\rev{To evaluate the expected Stetson J statistic, we used 100 mock AGN lightcurves from \cite{Fagin:2025:AGN_sims} observed with a daily cadence for 32 days at fiducial uncertainty of 5 mmags  to calculate a median Stetson \( J \) of 3. While this number is low, it also differs significantly from our observed sample, where only 10\% of our reliable matches to AGN have a Stetson \( J \) greater than 3. However, by changing the fiducial uncertainty to 0.03 mag, we find a similar simulated distribution of Stetson \( J \) as our observed sample.} 

\rev{Using the same mock AGN lightcurves, we also simulated full 10 year lightcurves with observations at LSST Wide Fast Deep (WFD) cadence from the v5.0.0\_10yrs baseline, incorporating Rubin errors using the Rubin Operations Simulator (OpSim) \citep{Reuter:2016:LSST_opsim}. The median r-band Stetson \( J \) over this baseline is 18.0, and the median Stetson \( J \) after one year is 13.0.} Thus the upcoming LSST survey will probe nearly the full range of variability timescales, and so similar analyses may be more informative in the future\rev{, including after the release of DR1}.

\subsection{Plots in optical-X-ray parameter space}

\begin{figure*}
    \centering
    \includegraphics[width=0.97\textwidth]{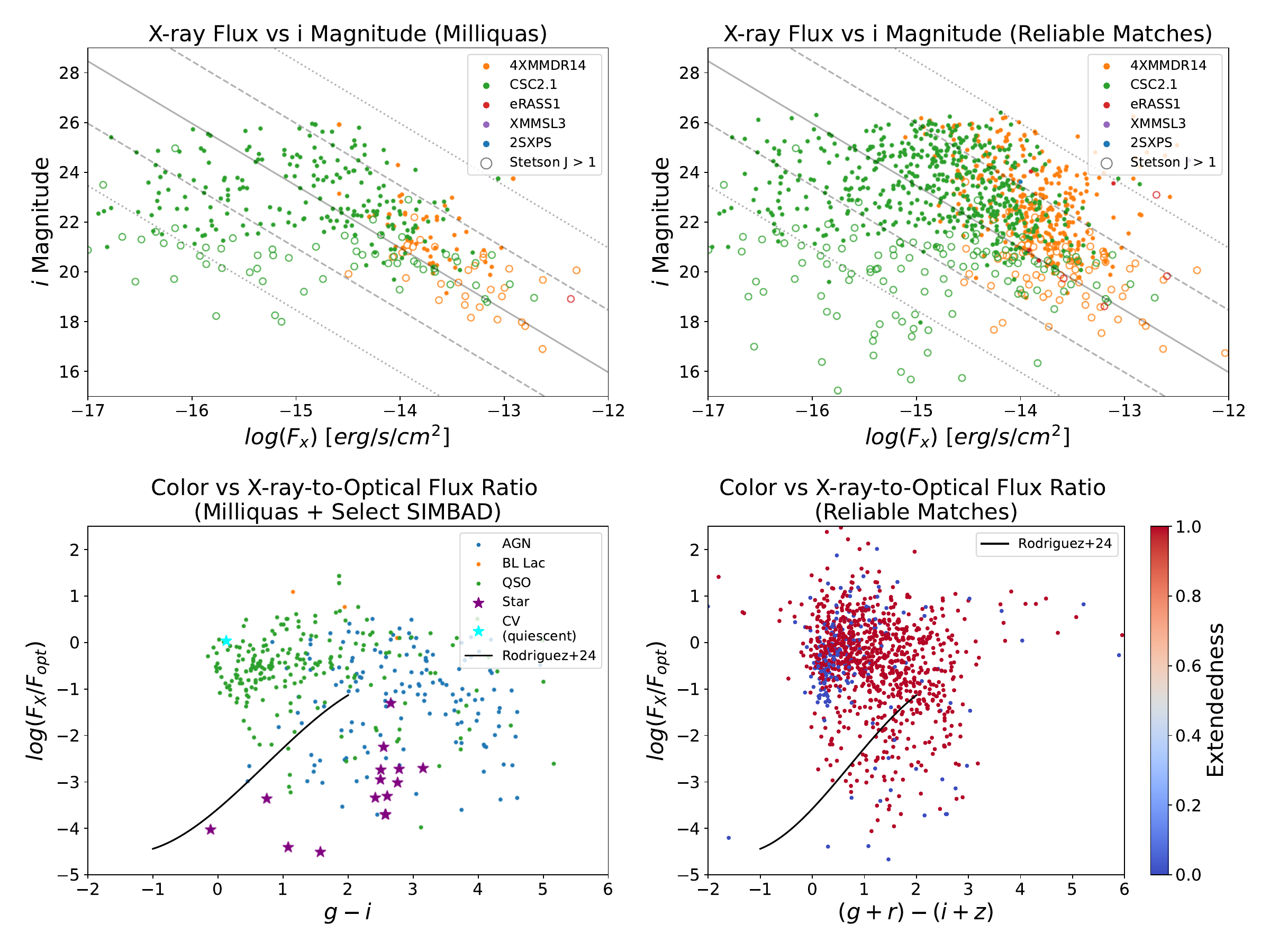}
    \caption{TOP: On the left, we plot the X-ray flux vs i-band magnitude, only for objects with matches to the Milliquas catalog. The solid grey line delineates $\log(F_{X}/F_{opt}) = 0$, and the dashed and dotted lines are at log flux ratios of $\pm 1$ and $\pm2$ respectively. Most of the X-ray bright galaxies fall between $\pm 1$, as expected for AGN. \rev{We expect many of the matches with $\log(F_{X}/F_{opt}) < -1$ to be galaxies.} The right plot shows all matches with reliability $>90\%$, a total of 1344 \Object{}s, on the same axes. \rev{The open circles show matches with some optical variability as defined by Stetson \( J \); however, it appears there is a strong correlation between optical magnitude and Stetson \( J \) as mentioned previously.}
    BOTTOM: We plot the X-ray-to-optical flux ratio against color. The left plot contains matches to Milliquas, plus some stars from SIMBAD and Gaia, and one CV (CRTS J033349.8-282244). The right plot again shows all matches with reliability $>90\%$ on the same axes. Extendedness measures whether the matched DP1 \Object{} is point-like (0) or extended (1). We use color transformations as defined in our Equation \ref{eq:gaia_lsst_color} to plot the empirical cut from \cite{Rodriguez:2024:X-ray_MS}.}
    \label{fig:fx_to_fopt}
\end{figure*}

A full classification of all the X-ray sources with DP1 counterparts is beyond the scope of this paper. However, we take advantage of our crossmatch to the Milliquas catalog and our SIMBAD queries to show where a sample of objects of known object type lie in our optical-X-ray parameter space. In Figure \ref{fig:fx_to_fopt}, we make two sets of two plots each. On the left, we plot ``control samples'' of classified objects. On the right, we plot all matched DP1 \Object{}s that have a reliability greater than 90\%.

The top left subplot shows our 444 \Object{} sample of matches to the Milliquas catalog in the i-magnitude / X-ray broadband flux plane, similar to figure 3 in \cite{Hasinger:2008:AGN_properties}. Our results are consistent, with most sources falling between the lines of constant X-ray to optical flux ratio of $\log(F_X/F_{opt}) \pm 1$\rev{, where $F_{opt}$ is the optical flux over all available \textit{ugrizy} bands}. For sources with fluxes between $10^{-14}$ and $10^{-12}\,{\rm erg}\,{\rm s}^{-1}\,{\rm cm}^{-2}$. Below $10^{-14}\,{\rm erg} {\rm s}^{-1} {\rm cm}^{-2}$ it appears some of the AGN in our sample have smaller $\log(F_X/F_{opt})$.  We also see much more scatter in the top right plot, even though we expect most of the sources to be extragalactic as well.

The bottom left subplot shows the log flux ratio $\log(F_X/F_{opt})$ against the color. We choose to plot $g-i$ color, and plot the empirical cut from \cite{Rodriguez:2024:X-ray_MS} onto our subplot using the following transformation:
\begin{equation}
    y = -0.079 + 1.184x + 0.2218x^2 - 0.1013x^3
    \label{eq:gaia_lsst_color}
\end{equation}
\new{\rev{w}here y is LSST $g-i$ color and x is Gaia $BP-RP$, for x between approximately $-1$ and $2$ \citep{Gwyn:2024:gaia_lsst_bands}.} This cut and diagram, termed the X-ray Main Sequence, is a useful tool for separating Galactic X-ray sources into stellar and compact object regions. Although the vast majority of our sources are extragalactic, we plot our Milliquas matches plus a handful of known Galactic stars and one cataclysmic variable (CRTS J033349.8-282244). We note that this CV was actually in outburst (see Figure \ref{fig:cv_lightcurve}), and we calculate the optical flux and color in quiescence by selecting the forced photometry before and after the outburst. The known stars fall on the opposite side of the cut as the CV. We also see that the quasi-stellar objects (QSOs) are much bluer and have a higher flux ratio than the AGN.

On the bottom right, we again plot the matched DP1 \Object{}s that have a reliability greater than 90\% in the same parameter space. On color, we have the \texttt{refExtendedness}, a measure of whether the DP1 \Object{} is point-like (0) or extended (1). These classifications are roughly meant to correspond to star-like and galactic sources. However, from our sample of known sources, we find that this is not always dependable, as QSOs are classified as point-like about 1/3 of the time (77/221). We also notice that our CV is classified as an``extended'' source as well. For future followup, we would begin by selecting the set of sources in the upper right of the plot with point-like \texttt{refExtendedness} and in fields other than E-CDF-S and EDF-S for investigation as potential accreting systems. 

\subsection{\diffIm{}s and crossmatch to the \DiaObject{} catalog} \label{sec:dia}
DP1 also contains catalogs of detections by the Difference Image Analysis pipeline (DIA) in the \diffIm{}s. Detections on individual \diffIm{}s are known as \DiaSource{}s and the difference image analysis pipeline associates \DiaSource{}s together to create a catalog of \DiaObject{}s, the equivalent to the \Object{} table for difference image analysis. We expect transients and variables to show up in the \DiaObject{} catalog and thus this is another way to assess whether an \Object{} is variable. Similar to the \ForcedSource{} catalog, DP1 also includes a \ForcedSourceOnDiaObject{} \citep{ForcedSourceOnDiaObject} catalog containing forced photometry on the \visitIm{}s at the location of each \DiaObject{}. Additionally, the \DiaSource{} catalog contains all photometry of each \DiaSource{} that is associated with a given \DiaObject{}. 

We crossmatch the locations of all the X-ray sources to the \DiaObject{} catalog and find 2276 candidate \DiaObject{}s for 1067 unique X-ray sources. 815 X-ray sources have counterparts in both the \Object{} and \DiaObject{} catalogs, although we cannot confirm that the \DiaObject{} and \Object{} are the same astronomical object for all 815 sources. Thus, of the 3830 X-ray sources in the combined X-ray catalog, 2566 have at least one match between the \Object{} and \DiaObject{} tables. We note that many X-ray sources in the 47 Tuc and Fornax dSph fields only have matches to the \DiaObject{} catalog. 

Since there is no \diffIm{} equivalent to a co-added image, we calculate the error-weighted mean magnitude in each band of every \DiaObject{} using the \texttt{psfMag} and \texttt{psfMagErr} columns in the \ForcedSourceOnDiaObject{} table as the per-band photometry of each \DiaObject{}.

\section{Conclusions} \label{sec:concl}
We performed a crossmatch between a consolidated catalog of all objects in the DP1 fields in five X-ray catalogs to DP1 \texttt{Objects} and \texttt{diaObjects}. Our results demonstrate that we are able to find many candidate counterparts with high reliability. Many of the candidate counterparts are fainter than Gaia, PanSTARRs, and SDSS limiting magnitudes. We find that 2566 out of 3830, or roughly 67\% of our X-ray sources have a match in one of the DP1 catalogs. After examining the X-ray to optical flux ratios of these matches, we do not follow up any of these sources at the present. A list of X-ray sources with optical non-detections, which may also be of interest for future followup, can be found in the linked GitHub. 

Many previous studies with deeper X-ray to optical crossmatches focus on a specific field such as E-CDF-S \citep{Luo:2017:ECDFS}, the Lockman Hole \citep{Salvato:2018:NWAY, Bykov:2022:eROSITA_Lockman_ML}, or the galactic bulge \citep{Weveres:2016:GBS_xmatch}. The LSST DP1 dataset contains six fields with very different density, observing depth, and coverage by X-ray catalogs, and we attempted to apply the same crossmatching procedure to each field. We find that many of our crossmatches are reliable in fields that are well observed by X-ray missions, i.e. the E-CDF-S field \rev{and that in this field, the percentage of X-ray objects with a reliable counterpart does not seem to depend on its X-ray flux.} In the non-crowded fields where we expect the completeness of the \Object{} catalog to be high, we find that at \rev{roughly 20\% of our matches met our criteria for reliability when the X-ray flux exceeds $10^{-13}\,{\rm erg} {\rm s}^{-1} {\rm cm}^{-2}$, and steadily drops off to zero for the faintest sources in these fields, at $10^{-15}\,{\rm erg} {\rm s}^{-1} {\rm cm}^{-2}$. Observations in these fields are primarily from eROSITA, and are limited by the sensitivity and the positional error of the eRASS1 data.}. In the crowded fields 47 Tuc and Fornax dSph, the majority of matches are found only on \texttt{diaObjects}, which bypass the deblending on co-add image step of the pipeline. We expect that it will be feasible to build up the large datasets of reliable optical-X-ray matches across the LSST all-sky survey to extract the local magnitude and sky density by magnitude distributions needed to implement and train more sophisticated crossmatching models but that this will require much larger survey area than the current DP1 fields. \rev{For example, \cite{Roster:2025:Euclid_Xray} use an X-ray emitter prior in their crossmatch between X-ray sources and Euclid data based on a machine learning algorithm.}

The variability quantified by the Stetson \( J \) index does not appear to \rev{strongly} correlate with the reliability of the crossmatch \rev{in the DP1 data}, and it remains inconclusive whether increased optical variability can indicate a higher likelihood counterpart, despite our \rev{expectation} that should be the case. As the Rubin \rev{data processing continues to be improved we will check back to see if this effect persists. Additionally, the continuous cadence of the commissioning observations mean the Stetson \( J \)s are most sensitive to variability on timescales of seconds to days, since the DP1 observations were on these timescales. This differs from much of the expected longer variability timescales of most extragalactic X-ray counterparts. However, the upcoming LSST full survey will probe nearly the full range of variability timescales and we plan to redo this analysis in the future once the survey is underway.}

Overall, while challenges remain, the DP1 data shows promise for X-ray optical association. Continued improvements to the Rubin Science Pipelines and in-kind programs, particularly those dedicated to crowded-field stellar photometry, are expected to deliver even better results in future data previews and data releases. These broader datasets should permit the utilization of more advanced statistical and machine learning based crossmatching techniques that can that will enable comprehensive classifications of X-ray sources and searches for rare and exotic new objects.

\section*{Acknowledgments}

YDW and ECB acknowledge support from the Vera C.\ Rubin Observatory, which is supported in part by the National Science Foundation through Cooperative Agreements AST-1258333 and AST-2241526 and Cooperative Support Agreements AST-1202910 and 2211468 managed by the Association of Universities for Research in Astronomy (AURA), and the Department of Energy under Contract No. DE-AC02-76SF00515 with the SLAC National Accelerator Laboratory managed by Stanford University. Additional Rubin Observatory funding comes from private donations, grants to universities, and in-kind support from LSST-DA Institutional Members.

YDW and ECB acknowledge support from the DiRAC Institute in the Department of Astronomy at the University of Washington. 
The DiRAC Institute is supported through generous gifts from the Charles and Lisa Simonyi Fund for Arts and Sciences, Janet and Lloyd Frink, and the Washington Research Foundation.

Support was provided by Schmidt Sciences, LLC. for S.~Campos, N.~Caplar, and K.~Malanchev. PG acknowledges support from the Science and Technology Facilities Council (STFC) and is a Royal Society Leverhulme Trust Senior Research Fellow. The Center for Computational Astrophysics at the Flatiron Institute is supported by the Simons Foundation. 

This research has made use of data obtained from the \XMM slew survey source catalogue compiled by the \XMM Survey Science Centre consortium in collaboration with members of the \XMM SOC and the EPIC consortium.

This research has made use of data obtained from the 4XMM \XMM serendipitous source catalogue compiled by the \XMM Survey Science Centre consortium.

This research has made use of data obtained from the Chandra Source Catalog, provided by the Chandra X-ray Center (CXC).

This work made use of data supplied by the UK \Swift Science Data Centre at the University of Leicester.

This work is based on data from \eROSITA, the soft X-ray instrument aboard SRG, a joint Russian-German science mission supported by the Russian Space Agency (Roskosmos), in the interests of the Russian Academy of Sciences represented by its Space Research Institute (IKI), and the Deutsches Zentrum für Luft- und Raumfahrt (DLR). The SRG spacecraft was built by Lavochkin Association (NPOL) and its subcontractors, and is operated by NPOL with support from the Max Planck Institute for Extraterrestrial Physics (MPE). The development and construction of the \eROSITA X-ray instrument was led by MPE, with contributions from the Dr. Karl Remeis Observatory Bamberg \& ECAP (FAU Erlangen-Nuernberg), the University of Hamburg Observatory, the Leibniz Institute for Astrophysics Potsdam (AIP), and the Institute for Astronomy and Astrophysics of the University of Tübingen, with the support of DLR and the Max Planck Society. The Argelander Institute for Astronomy of the University of Bonn and the Ludwig Maximilians Universität Munich also participated in the science preparation for \eROSITA.

\facilities{Rubin:Simonyi (LSSTComCam), USDAC, USDF, XMM-Newton, Chandra, Swift, eROSITA, Gaia}

\software{\texttt{LSST Science Pipelines} \citep{PSTN-019}, \texttt{Rubin Data Butler}\citep{Jenness:2022:butler}, \texttt{Astropy} \citep{Astropy2013, Astropy2018, Astropy2022}, \texttt{Astroquery} \citep{Ginsburg:2019:Astroquery}, \texttt{Matplotlib} \citep{Hunter2007}, \texttt{NumPy} \citep{vanderwalt2011, harris2020}, \texttt{LSDB} \citep{Caplar:2025:LSDB}.}

\bibliographystyle{aasjournal}

\bibliography{arxiv_ms}

@TechReport{RDO-013,
    author = "Blum, Robert and the Rubin Operations Team",
    title = "{Vera C. Rubin Observatory Data Policy}",
    year = "2020",
    month = "September",
    institution = "Vera C. Rubin Observatory",
    url = "https://ls.st/RDO-013",
    type = "{Data Management Operations Controlled Document}",
    number = "RDO-013",
    handle = "RDO-013"
}

@misc{Rubin:DP1,
    author = {{NSF-DOE Vera C. Rubin Observatory}},
year = 2025,
title = {{The Vera C. Rubin Observatory Data Preview 1}},
doi = {10.71929/rubin/2570536},
url = {https://doi.org/10.71929/rubin/2570536},
}

@ARTICLE{Sutherland:1992:LR,
       author = {{Sutherland}, Will and {Saunders}, Will},
        title = "{On the likelihood ratio for source identification.}",
      journal = {\mnras},
         year = 1992,
        month = dec,
       volume = {259},
        pages = {413-420},
          doi = {10.1093/mnras/259.3.413},
       adsurl = {https://ui.adsabs.harvard.edu/abs/1992MNRAS.259..413S}
}

@ARTICLE{Richter:1975:LR,
       author = {{Richter}, G.~A.},
        title = "{Search for Optical Identifications in the 5C3 Radio Survey. II. Statistical Treatment and Results}",
      journal = {Astronomische Nachrichten},
         year = 1975,
        month = jan,
       volume = {296},
       number = {2},
        pages = {65},
          doi = {10.1002/asna.19752960203},
       adsurl = {https://ui.adsabs.harvard.edu/abs/1975AN....296...65R}
}

@ARTICLE{Webb:2020:4XMMDR14,
       author = {{Webb}, N.~A. and {Coriat}, M. and {Traulsen}, I. and {Ballet}, J. and {Motch}, C. and {Carrera}, F.~J. and {Koliopanos}, F. and {Authier}, J. and {de la Calle}, I. and {Ceballos}, M.~T. and {Colomo}, E. and {Chuard}, D. and {Freyberg}, M. and {Garcia}, T. and {Kolehmainen}, M. and {Lamer}, G. and {Lin}, D. and {Maggi}, P. and {Michel}, L. and {Page}, C.~G. and {Page}, M.~J. and {Perea-Calderon}, J.~V. and {Pineau}, F. -X. and {Rodriguez}, P. and {Rosen}, S.~R. and {Santos Lleo}, M. and {Saxton}, R.~D. and {Schwope}, A. and {Tom{\'a}s}, L. and {Watson}, M.~G. and {Zakardjian}, A.},
        title = "{The XMM-Newton serendipitous survey. IX. The fourth XMM-Newton serendipitous source catalogue}",
      journal = {\aap},
         year = 2020,
        month = sep,
       volume = {641},
          eid = {A136},
        pages = {A136},
          doi = {10.1051/0004-6361/201937353},
archivePrefix = {arXiv},
       eprint = {2007.02899},
 primaryClass = {astro-ph.HE},
       adsurl = {https://ui.adsabs.harvard.edu/abs/2020A&A...641A.136W}
}

@ARTICLE{Saxton:2008:XMMSL,
       author = {{Saxton}, R.~D. and {Read}, A.~M. and {Esquej}, P. and {Freyberg}, M.~J. and {Altieri}, B. and {Bermejo}, D.},
        title = "{The first XMM-Newton slew survey catalogue: XMMSL1}",
      journal = {\aap},
         year = 2008,
        month = mar,
       volume = {480},
       number = {2},
        pages = {611-622},
          doi = {10.1051/0004-6361:20079193},
archivePrefix = {arXiv},
       eprint = {0801.3732},
 primaryClass = {astro-ph},
       adsurl = {https://ui.adsabs.harvard.edu/abs/2008A&A...480..611S}
}

@ARTICLE{Evans:2024:CSC2.1,
       author = {{Evans}, Ian N. and {Evans}, Janet D. and {Mart{\'\i}nez-Galarza}, J. Rafael and {Miller}, Joseph B. and {Primini}, Francis A. and {Azadi}, Mojegan and {Burke}, Douglas J. and {Civano}, Francesca M. and {D'Abrusco}, Raffaele and {Fabbiano}, Giuseppina and {Graessle}, Dale E. and {Grier}, John D. and {Houck}, John C. and {Lauer}, Jennifer and {McCollough}, Michael L. and {Nowak}, Michael A. and {Plummer}, David A. and {Rots}, Arnold H. and {Siemiginowska}, Aneta and {Tibbetts}, Michael S.},
        title = "{The Chandra Source Catalog Release 2 Series}",
      journal = {\apjs},
         year = 2024,
        month = oct,
       volume = {274},
       number = {2},
          eid = {22},
        pages = {22},
          doi = {10.3847/1538-4365/ad6319},
archivePrefix = {arXiv},
       eprint = {2407.10799},
 primaryClass = {astro-ph.HE},
       adsurl = {https://ui.adsabs.harvard.edu/abs/2024ApJS..274...22E}
}

@ARTICLE{Evans:2020:2SXPS,
       author = {{Evans}, P.~A. and {Page}, K.~L. and {Osborne}, J.~P. and {Beardmore}, A.~P. and {Willingale}, R. and {Burrows}, D.~N. and {Kennea}, J.~A. and {Perri}, M. and {Capalbi}, M. and {Tagliaferri}, G. and {Cenko}, S.~B.},
        title = "{2SXPS: An Improved and Expanded Swift X-Ray Telescope Point-source Catalog}",
      journal = {\apjs},
         year = 2020,
        month = apr,
       volume = {247},
       number = {2},
          eid = {54},
        pages = {54},
          doi = {10.3847/1538-4365/ab7db9},
archivePrefix = {arXiv},
       eprint = {1911.11710},
 primaryClass = {astro-ph.IM},
       adsurl = {https://ui.adsabs.harvard.edu/abs/2020ApJS..247...54E}
}

@ARTICLE{Merloni:2024:eRASS1,
       author = {{Merloni}, A. and {Lamer}, G. and {Liu}, T. and {Ramos-Ceja}, M.~E. and {Brunner}, H. and {Bulbul}, E. and {Dennerl}, K. and {Doroshenko}, V. and {Freyberg}, M.~J. and {Friedrich}, S. and {Gatuzz}, E. and {Georgakakis}, A. and {Haberl}, F. and {Igo}, Z. and {Kreykenbohm}, I. and {Liu}, A. and {Maitra}, C. and {Malyali}, A. and {Mayer}, M.~G.~F. and {Nandra}, K. and {Predehl}, P. and {Robrade}, J. and {Salvato}, M. and {Sanders}, J.~S. and {Stewart}, I. and {Tub{\'\i}n-Arenas}, D. and {Weber}, P. and {Wilms}, J. and {Arcodia}, R. and {Artis}, E. and {Aschersleben}, J. and {Avakyan}, A. and {Aydar}, C. and {Bahar}, Y.~E. and {Balzer}, F. and {Becker}, W. and {Berger}, K. and {Boller}, T. and {Bornemann}, W. and {Br{\"u}ggen}, M. and {Brusa}, M. and {Buchner}, J. and {Burwitz}, V. and {Camilloni}, F. and {Clerc}, N. and {Comparat}, J. and {Coutinho}, D. and {Czesla}, S. and {Dannhauer}, S.~M. and {Dauner}, L. and {Dauser}, T. and {Dietl}, J. and {Dolag}, K. and {Dwelly}, T. and {Egg}, K. and {Ehl}, E. and {Freund}, S. and {Friedrich}, P. and {Gaida}, R. and {Garrel}, C. and {Ghirardini}, V. and {Gokus}, A. and {Gr{\"u}nwald}, G. and {Grandis}, S. and {Grotova}, I. and {Gruen}, D. and {Gueguen}, A. and {H{\"a}mmerich}, S. and {Hamaus}, N. and {Hasinger}, G. and {Haubner}, K. and {Homan}, D. and {Ider Chitham}, J. and {Joseph}, W.~M. and {Joyce}, A. and {K{\"o}nig}, O. and {Kaltenbrunner}, D.~M. and {Khokhriakova}, A. and {Kink}, W. and {Kirsch}, C. and {Kluge}, M. and {Knies}, J. and {Krippendorf}, S. and {Krumpe}, M. and {Kurpas}, J. and {Li}, P. and {Liu}, Z. and {Locatelli}, N. and {Lorenz}, M. and {M{\"u}ller}, S. and {Magaudda}, E. and {Mannes}, C. and {McCall}, H. and {Meidinger}, N. and {Michailidis}, M. and {Migkas}, K. and {Mu{\~n}oz-Giraldo}, D. and {Musiimenta}, B. and {Nguyen-Dang}, N.~T. and {Ni}, Q. and {Olechowska}, A. and {Ota}, N. and {Pacaud}, F. and {Pasini}, T. and {Perinati}, E. and {Pires}, A.~M. and {Pommranz}, C. and {Ponti}, G. and {Poppenhaeger}, K. and {P{\"u}hlhofer}, G. and {Rau}, A. and {Reh}, M. and {Reiprich}, T.~H. and {Roster}, W. and {Saeedi}, S. and {Santangelo}, A. and {Sasaki}, M. and {Schmitt}, J. and {Schneider}, P.~C. and {Schrabback}, T. and {Schuster}, N. and {Schwope}, A. and {Seppi}, R. and {Serim}, M.~M. and {Shreeram}, S. and {Sokolova-Lapa}, E. and {Starck}, H. and {Stelzer}, B. and {Stierhof}, J. and {Suleimanov}, V. and {Tenzer}, C. and {Traulsen}, I. and {Tr{\"u}mper}, J. and {Tsuge}, K. and {Urrutia}, T. and {Veronica}, A. and {Waddell}, S.~G.~H. and {Willer}, R. and {Wolf}, J. and {Yeung}, M.~C.~H. and {Zainab}, A. and {Zangrandi}, F. and {Zhang}, X. and {Zhang}, Y. and {Zheng}, X.},
        title = "{The SRG/eROSITA all-sky survey. First X-ray catalogues and data release of the western Galactic hemisphere}",
      journal = {\aap},
         year = 2024,
        month = feb,
       volume = {682},
          eid = {A34},
        pages = {A34},
          doi = {10.1051/0004-6361/202347165},
archivePrefix = {arXiv},
       eprint = {2401.17274},
 primaryClass = {astro-ph.HE},
       adsurl = {https://ui.adsabs.harvard.edu/abs/2024A&A...682A..34M}
}

@ARTICLE{Luo:2017:ECDFS,
       author = {{Luo}, B. and {Brandt}, W.~N. and {Xue}, Y.~Q. and {Lehmer}, B. and {Alexander}, D.~M. and {Bauer}, F.~E. and {Vito}, F. and {Yang}, G. and {Basu-Zych}, A.~R. and {Comastri}, A. and {Gilli}, R. and {Gu}, Q. -S. and {Hornschemeier}, A.~E. and {Koekemoer}, A. and {Liu}, T. and {Mainieri}, V. and {Paolillo}, M. and {Ranalli}, P. and {Rosati}, P. and {Schneider}, D.~P. and {Shemmer}, O. and {Smail}, I. and {Sun}, M. and {Tozzi}, P. and {Vignali}, C. and {Wang}, J. -X.},
        title = "{The Chandra Deep Field-South Survey: 7 Ms Source Catalogs}",
      journal = {\apjs},
         year = 2017,
        month = jan,
       volume = {228},
       number = {1},
          eid = {2},
        pages = {2},
          doi = {10.3847/1538-4365/228/1/2},
archivePrefix = {arXiv},
       eprint = {1611.03501},
 primaryClass = {astro-ph.GA},
       adsurl = {https://ui.adsabs.harvard.edu/abs/2017ApJS..228....2L}
}

@ARTICLE{Rodriguez:2024:X-ray_MS,
       author = {{Rodriguez}, Antonio C.},
        title = "{From Active Stars to Black Holes: A Discovery Tool for Galactic X-Ray Sources}",
      journal = {\pasp},
         year = 2024,
        month = may,
       volume = {136},
       number = {5},
          eid = {054201},
        pages = {054201},
          doi = {10.1088/1538-3873/ad357c},
archivePrefix = {arXiv},
       eprint = {2401.09537},
 primaryClass = {astro-ph.HE},
       adsurl = {https://ui.adsabs.harvard.edu/abs/2024PASP..136e4201R}
}

@ARTICLE{Weveres:2016:GBS_xmatch,
       author = {{Wevers}, T. and {Hodgkin}, S.~T. and {Jonker}, P.~G. and {Bassa}, C. and {Nelemans}, G. and {van Grunsven}, T. and {Gonzalez-Solares}, E.~A. and {Torres}, M.~A.~P. and {Heinke}, C. and {Steeghs}, D. and {Maccarone}, T.~J. and {Britt}, C. and {Hynes}, R.~I. and {Johnson}, C. and {Wu}, Jianfeng},
        title = "{The Chandra Galactic Bulge Survey: optical catalogue and point-source counterparts to X-ray sources}",
      journal = {\mnras},
         year = 2016,
        month = jun,
       volume = {458},
       number = {4},
        pages = {4530-4546},
          doi = {10.1093/mnras/stw643},
archivePrefix = {arXiv},
       eprint = {1605.02741},
 primaryClass = {astro-ph.HE},
       adsurl = {https://ui.adsabs.harvard.edu/abs/2016MNRAS.458.4530W}
}

@ARTICLE{Ivezic:2019:LSST,
       author = {{Ivezi{\'c}}, {\v{Z}}eljko and {Kahn}, Steven M. and {Tyson}, J. Anthony and {Abel}, Bob and {Acosta}, Emily and {Allsman}, Robyn and {Alonso}, David and {AlSayyad}, Yusra and {Anderson}, Scott F. and {Andrew}, John and {Angel}, James Roger P. and {Angeli}, George Z. and {Ansari}, Reza and {Antilogus}, Pierre and {Araujo}, Constanza and {Armstrong}, Robert and {Arndt}, Kirk T. and {Astier}, Pierre and {Aubourg}, {\'E}ric and {Auza}, Nicole and {Axelrod}, Tim S. and {Bard}, Deborah J. and {Barr}, Jeff D. and {Barrau}, Aurelian and {Bartlett}, James G. and {Bauer}, Amanda E. and {Bauman}, Brian J. and {Baumont}, Sylvain and {Bechtol}, Ellen and {Bechtol}, Keith and {Becker}, Andrew C. and {Becla}, Jacek and {Beldica}, Cristina and {Bellavia}, Steve and {Bianco}, Federica B. and {Biswas}, Rahul and {Blanc}, Guillaume and {Blazek}, Jonathan and {Blandford}, Roger D. and {Bloom}, Josh S. and {Bogart}, Joanne and {Bond}, Tim W. and {Booth}, Michael T. and {Borgland}, Anders W. and {Borne}, Kirk and {Bosch}, James F. and {Boutigny}, Dominique and {Brackett}, Craig A. and {Bradshaw}, Andrew and {Brandt}, William Nielsen and {Brown}, Michael E. and {Bullock}, James S. and {Burchat}, Patricia and {Burke}, David L. and {Cagnoli}, Gianpietro and {Calabrese}, Daniel and {Callahan}, Shawn and {Callen}, Alice L. and {Carlin}, Jeffrey L. and {Carlson}, Erin L. and {Chandrasekharan}, Srinivasan and {Charles-Emerson}, Glenaver and {Chesley}, Steve and {Cheu}, Elliott C. and {Chiang}, Hsin-Fang and {Chiang}, James and {Chirino}, Carol and {Chow}, Derek and {Ciardi}, David R. and {Claver}, Charles F. and {Cohen-Tanugi}, Johann and {Cockrum}, Joseph J. and {Coles}, Rebecca and {Connolly}, Andrew J. and {Cook}, Kem H. and {Cooray}, Asantha and {Covey}, Kevin R. and {Cribbs}, Chris and {Cui}, Wei and {Cutri}, Roc and {Daly}, Philip N. and {Daniel}, Scott F. and {Daruich}, Felipe and {Daubard}, Guillaume and {Daues}, Greg and {Dawson}, William and {Delgado}, Francisco and {Dellapenna}, Alfred and {de Peyster}, Robert and {de Val-Borro}, Miguel and {Digel}, Seth W. and {Doherty}, Peter and {Dubois}, Richard and {Dubois-Felsmann}, Gregory P. and {Durech}, Josef and {Economou}, Frossie and {Eifler}, Tim and {Eracleous}, Michael and {Emmons}, Benjamin L. and {Fausti Neto}, Angelo and {Ferguson}, Henry and {Figueroa}, Enrique and {Fisher-Levine}, Merlin and {Focke}, Warren and {Foss}, Michael D. and {Frank}, James and {Freemon}, Michael D. and {Gangler}, Emmanuel and {Gawiser}, Eric and {Geary}, John C. and {Gee}, Perry and {Geha}, Marla and {Gessner}, Charles J.~B. and {Gibson}, Robert R. and {Gilmore}, D. Kirk and {Glanzman}, Thomas and {Glick}, William and {Goldina}, Tatiana and {Goldstein}, Daniel A. and {Goodenow}, Iain and {Graham}, Melissa L. and {Gressler}, William J. and {Gris}, Philippe and {Guy}, Leanne P. and {Guyonnet}, Augustin and {Haller}, Gunther and {Harris}, Ron and {Hascall}, Patrick A. and {Haupt}, Justine and {Hernandez}, Fabio and {Herrmann}, Sven and {Hileman}, Edward and {Hoblitt}, Joshua and {Hodgson}, John A. and {Hogan}, Craig and {Howard}, James D. and {Huang}, Dajun and {Huffer}, Michael E. and {Ingraham}, Patrick and {Innes}, Walter R. and {Jacoby}, Suzanne H. and {Jain}, Bhuvnesh and {Jammes}, Fabrice and {Jee}, M. James and {Jenness}, Tim and {Jernigan}, Garrett and {Jevremovi{\'c}}, Darko and {Johns}, Kenneth and {Johnson}, Anthony S. and {Johnson}, Margaret W.~G. and {Jones}, R. Lynne and {Juramy-Gilles}, Claire and {Juri{\'c}}, Mario and {Kalirai}, Jason S. and {Kallivayalil}, Nitya J. and {Kalmbach}, Bryce and {Kantor}, Jeffrey P. and {Karst}, Pierre and {Kasliwal}, Mansi M. and {Kelly}, Heather and {Kessler}, Richard and {Kinnison}, Veronica and {Kirkby}, David and {Knox}, Lloyd and {Kotov}, Ivan V. and {Krabbendam}, Victor L. and {Krughoff}, K. Simon and {Kub{\'a}nek}, Petr and {Kuczewski}, John and {Kulkarni}, Shri and {Ku}, John and {Kurita}, Nadine R. and {Lage}, Craig S. and {Lambert}, Ron and {Lange}, Travis and {Langton}, J. Brian and {Le Guillou}, Laurent and {Levine}, Deborah and {Liang}, Ming and {Lim}, Kian-Tat and {Lintott}, Chris J. and {Long}, Kevin E. and {Lopez}, Margaux and {Lotz}, Paul J. and {Lupton}, Robert H. and {Lust}, Nate B. and {MacArthur}, Lauren A. and {Mahabal}, Ashish and {Mandelbaum}, Rachel and {Markiewicz}, Thomas W. and {Marsh}, Darren S. and {Marshall}, Philip J. and {Marshall}, Stuart and {May}, Morgan and {McKercher}, Robert and {McQueen}, Michelle and {Meyers}, Joshua and {Migliore}, Myriam and {Miller}, Michelle and {Mills}, David J.},
        title = "{LSST: From Science Drivers to Reference Design and Anticipated Data Products}",
      journal = {\apj},
         year = 2019,
        month = mar,
       volume = {873},
       number = {2},
          eid = {111},
        pages = {111},
          doi = {10.3847/1538-4357/ab042c},
archivePrefix = {arXiv},
       eprint = {0805.2366},
 primaryClass = {astro-ph},
       adsurl = {https://ui.adsabs.harvard.edu/abs/2019ApJ...873..111I}
}

@ARTICLE{Caplar:2025:LSDB,
    author        = {{Caplar}, Neven and {Beebe}, Wilson and {Branton}, Doug and {Campos}, Sandro and {Connolly}, Andrew and {DeLucchi}, Melissa and {Jones}, Derek and {Juric}, Mario and {Kubica}, Jeremy and {Malanchev}, Konstantin and {Mandelbaum}, Rachel and {McGuire}, Sean},
    title         = "{Using LSDB to enable large-scale catalog distribution, cross-matching, and analytics}",
    journal       = {arXiv e-prints},
    year          = 2025,
    month         = jan,
    eid           = {arXiv:2501.02103},
    pages         = {arXiv:2501.02103},
    doi           = {10.48550/arXiv.2501.02103},
    archivePrefix = {arXiv},
    eprint        = {2501.02103},
    primaryClass  = {astro-ph.IM},
    adsurl        = {https://ui.adsabs.harvard.edu/abs/2025arXiv250102103C}
}

@ARTICLE{Sutherland:1992:LikelihoodRatio,
       author = {{Sutherland}, Will and {Saunders}, Will},
        title = "{On the likelihood ratio for source identification.}",
      journal = {\mnras},
         year = 1992,
        month = dec,
       volume = {259},
        pages = {413-420},
          doi = {10.1093/mnras/259.3.413},
       adsurl = {https://ui.adsabs.harvard.edu/abs/1992MNRAS.259..413S}
}

@ARTICLE{Salvato:2018:NWAY,
       author = {{Salvato}, M. and {Buchner}, J. and {Budav{\'a}ri}, T. and {Dwelly}, T. and {Merloni}, A. and {Brusa}, M. and {Rau}, A. and {Fotopoulou}, S. and {Nandra}, K.},
        title = "{Finding counterparts for all-sky X-ray surveys with NWAY: a Bayesian algorithm for cross-matching multiple catalogues}",
      journal = {\mnras},
         year = 2018,
        month = feb,
       volume = {473},
       number = {4},
        pages = {4937-4955},
          doi = {10.1093/mnras/stx2651},
archivePrefix = {arXiv},
       eprint = {1705.10711},
 primaryClass = {astro-ph.GA},
       adsurl = {https://ui.adsabs.harvard.edu/abs/2018MNRAS.473.4937S}
}

@ARTICLE{Bykov:2022:eROSITA_Lockman_ML,
       author = {{Bykov}, S.~D. and {Belvedersky}, M.~I. and {Gilfanov}, M.~R.},
        title = "{Optical Cross-Match of SRG/eROSITA X-ray Sources Using the Deep Lockman Hole Survey as an Example}",
      journal = {Astronomy Letters},
         year = 2022,
        month = nov,
       volume = {48},
       number = {11},
        pages = {653-664},
          doi = {10.1134/S1063773722110044},
archivePrefix = {arXiv},
       eprint = {2302.13689},
 primaryClass = {astro-ph.HE},
       adsurl = {https://ui.adsabs.harvard.edu/abs/2022AstL...48..653B}
}

@ARTICLE{Flesch:2023:Milliquas,
       author = {{Flesch}, Eric Wim},
        title = "{The Million Quasars (Milliquas) Catalogue, v8}",
      journal = {The Open Journal of Astrophysics},
         year = 2023,
        month = dec,
       volume = {6},
          eid = {49},
        pages = {49},
          doi = {10.21105/astro.2308.01505},
archivePrefix = {arXiv},
       eprint = {2308.01505},
 primaryClass = {astro-ph.GA},
       adsurl = {https://ui.adsabs.harvard.edu/abs/2023OJAp....6E..49F}
}

@misc{DiaObject,
  doi = {10.71929/RUBIN/2570319},
  url = {https://www.osti.gov//servlets/purl/2570319},
  author = {{NSF-DOE Vera C. Rubin Observatory}},
  title = {Legacy Survey of Space and Time Data Preview 1: DiaObject searchable catalog},
  publisher = {NSF-DOE Vera C. Rubin Observatory},
  year = {2025}
}

@TechReport{PSTN-019,
    author = "{Rubin Observatory Science Pipelines Developers}",
    title = "{The LSST Science Pipelines Software: Optical Survey Pipeline Reduction and Analysis Environment}",
    institution = "{Vera C. Rubin Observatory}",
    year = "2025",
    month = "June",
    handle = "PSTN-019",
    type = "{Project Science Technical Note}",
    number = "PSTN-019",
    doi = "{10.71929/rubin/2570545}",
    url = "https://pstn-019.lsst.io/"
}

@TechReport{RTN-011,
    author = "Guy, Leanne P. and Bechtol, Keith and Bellm, Eric and Blum, Bob and Dubois-Felsmann, Gregory P. and Graham, Melissa L. and Ivezi\'{c}, {\v Z}eljko and Lupton, Robert H. and Marshall, Phil and Slater, Colin T. and Strauss., Michael",
    title = "{Rubin Observatory Plans for an Early Science Program}",
    institution = "{Vera C. Rubin Observatory}",
    year = "2025",
    month = "May",
    handle = "RTN-011",
    type = "{Technical Note}",
    number = "RTN-011",
    url = "https://rtn-011.lsst.io/"
}

@ARTICLE{Stetson:1996:StetsonJ,
       author = {{Stetson}, Peter B.},
        title = "{On the Automatic Determination of Light-Curve Parameters for Cepheid Variables}",
      journal = {\pasp},
         year = 1996,
        month = oct,
       volume = {108},
        pages = {851},
          doi = {10.1086/133808},
       adsurl = {https://ui.adsabs.harvard.edu/abs/1996PASP..108..851S}
}

@INPROCEEDINGS{Jenness:2022:butler,
       author = {{Jenness}, Tim and {Bosch}, James F. and {Salnikov}, Andrei and {Lust}, Nate B. and {Pease}, Nathan M. and {Gower}, Michelle and {Kowalik}, Mikolaj and {Dubois-Felsmann}, Gregory P. and {Mueller}, Fritz and {Schellart}, Pim},
        title = "{The Vera C. Rubin Observatory Data Butler and pipeline execution system}",
    booktitle = {Software and Cyberinfrastructure for Astronomy VII},
         year = 2022,
       series = {Society of Photo-Optical Instrumentation Engineers (SPIE) Conference Series},
       volume = {12189},
        month = aug,
          eid = {1218911},
        pages = {1218911},
          doi = {10.1117/12.2629569},
archivePrefix = {arXiv},
       eprint = {2206.14941},
 primaryClass = {astro-ph.IM},
       adsurl = {https://ui.adsabs.harvard.edu/abs/2022SPIE12189E..11J}
}

@ARTICLE{Astropy2022,
       author = {{Astropy Collaboration} and {Price-Whelan}, Adrian M. and {Lim}, Pey Lian and {Earl}, Nicholas and {Starkman}, Nathaniel and {Bradley}, Larry and {Shupe}, David L. and {Patil}, Aarya A. and {Corrales}, Lia and {Brasseur}, C.~E. and {N{\"o}the}, Maximilian and {Donath}, Axel and {Tollerud}, Erik and {Morris}, Brett M. and {Ginsburg}, Adam and {Vaher}, Eero and {Weaver}, Benjamin A. and {Tocknell}, James and {Jamieson}, William and {van Kerkwijk}, Marten H. and {Robitaille}, Thomas P. and {Merry}, Bruce and {Bachetti}, Matteo and {G{\"u}nther}, H. Moritz and {Aldcroft}, Thomas L. and {Alvarado-Montes}, Jaime A. and {Archibald}, Anne M. and {B{\'o}di}, Attila and {Bapat}, Shreyas and {Barentsen}, Geert and {Baz{\'a}n}, Juanjo and {Biswas}, Manish and {Boquien}, M{\'e}d{\'e}ric and {Burke}, D.~J. and {Cara}, Daria and {Cara}, Mihai and {Conroy}, Kyle E. and {Conseil}, Simon and {Craig}, Matthew W. and {Cross}, Robert M. and {Cruz}, Kelle L. and {D'Eugenio}, Francesco and {Dencheva}, Nadia and {Devillepoix}, Hadrien A.~R. and {Dietrich}, J{\"o}rg P. and {Eigenbrot}, Arthur Davis and {Erben}, Thomas and {Ferreira}, Leonardo and {Foreman-Mackey}, Daniel and {Fox}, Ryan and {Freij}, Nabil and {Garg}, Suyog and {Geda}, Robel and {Glattly}, Lauren and {Gondhalekar}, Yash and {Gordon}, Karl D. and {Grant}, David and {Greenfield}, Perry and {Groener}, Austen M. and {Guest}, Steve and {Gurovich}, Sebastian and {Handberg}, Rasmus and {Hart}, Akeem and {Hatfield-Dodds}, Zac and {Homeier}, Derek and {Hosseinzadeh}, Griffin and {Jenness}, Tim and {Jones}, Craig K. and {Joseph}, Prajwel and {Kalmbach}, J. Bryce and {Karamehmetoglu}, Emir and {Ka{\l}uszy{\'n}ski}, Miko{\l}aj and {Kelley}, Michael S.~P. and {Kern}, Nicholas and {Kerzendorf}, Wolfgang E. and {Koch}, Eric W. and {Kulumani}, Shankar and {Lee}, Antony and {Ly}, Chun and {Ma}, Zhiyuan and {MacBride}, Conor and {Maljaars}, Jakob M. and {Muna}, Demitri and {Murphy}, N.~A. and {Norman}, Henrik and {O'Steen}, Richard and {Oman}, Kyle A. and {Pacifici}, Camilla and {Pascual}, Sergio and {Pascual-Granado}, J. and {Patil}, Rohit R. and {Perren}, Gabriel I. and {Pickering}, Timothy E. and {Rastogi}, Tanuj and {Roulston}, Benjamin R. and {Ryan}, Daniel F. and {Rykoff}, Eli S. and {Sabater}, Jose and {Sakurikar}, Parikshit and {Salgado}, Jes{\'u}s and {Sanghi}, Aniket and {Saunders}, Nicholas and {Savchenko}, Volodymyr and {Schwardt}, Ludwig and {Seifert-Eckert}, Michael and {Shih}, Albert Y. and {Jain}, Anany Shrey and {Shukla}, Gyanendra and {Sick}, Jonathan and {Simpson}, Chris and {Singanamalla}, Sudheesh and {Singer}, Leo P. and {Singhal}, Jaladh and {Sinha}, Manodeep and {Sip{\H{o}}cz}, Brigitta M. and {Spitler}, Lee R. and {Stansby}, David and {Streicher}, Ole and {{\v{S}}umak}, Jani and {Swinbank}, John D. and {Taranu}, Dan S. and {Tewary}, Nikita and {Tremblay}, Grant R. and {de Val-Borro}, Miguel and {Van Kooten}, Samuel J. and {Vasovi{\'c}}, Zlatan and {Verma}, Shresth and {de Miranda Cardoso}, Jos{\'e} Vin{\'\i}cius and {Williams}, Peter K.~G. and {Wilson}, Tom J. and {Winkel}, Benjamin and {Wood-Vasey}, W.~M. and {Xue}, Rui and {Yoachim}, Peter and {Zhang}, Chen and {Zonca}, Andrea and {Astropy Project Contributors}},
        title = "{The Astropy Project: Sustaining and Growing a Community-oriented Open-source Project and the Latest Major Release (v5.0) of the Core Package}",
      journal = {\apj},
         year = 2022,
        month = aug,
       volume = {935},
       number = {2},
          eid = {167},
        pages = {167},
          doi = {10.3847/1538-4357/ac7c74},
archivePrefix = {arXiv},
       eprint = {2206.14220},
 primaryClass = {astro-ph.IM},
       adsurl = {https://ui.adsabs.harvard.edu/abs/2022ApJ...935..167A}
}

@ARTICLE{Astropy2018,
       author = {{Astropy Collaboration} and {Price-Whelan}, A.~M. and {Sip{\H{o}}cz}, B.~M. and {G{\"u}nther}, H.~M. and {Lim}, P.~L. and {Crawford}, S.~M. and {Conseil}, S. and {Shupe}, D.~L. and {Craig}, M.~W. and {Dencheva}, N. and {Ginsburg}, A. and {VanderPlas}, J.~T. and {Bradley}, L.~D. and {P{\'e}rez-Su{\'a}rez}, D. and {de Val-Borro}, M. and {Aldcroft}, T.~L. and {Cruz}, K.~L. and {Robitaille}, T.~P. and {Tollerud}, E.~J. and {Ardelean}, C. and {Babej}, T. and {Bach}, Y.~P. and {Bachetti}, M. and {Bakanov}, A.~V. and {Bamford}, S.~P. and {Barentsen}, G. and {Barmby}, P. and {Baumbach}, A. and {Berry}, K.~L. and {Biscani}, F. and {Boquien}, M. and {Bostroem}, K.~A. and {Bouma}, L.~G. and {Brammer}, G.~B. and {Bray}, E.~M. and {Breytenbach}, H. and {Buddelmeijer}, H. and {Burke}, D.~J. and {Calderone}, G. and {Cano Rodr{\'\i}guez}, J.~L. and {Cara}, M. and {Cardoso}, J.~V.~M. and {Cheedella}, S. and {Copin}, Y. and {Corrales}, L. and {Crichton}, D. and {D'Avella}, D. and {Deil}, C. and {Depagne}, {\'E}. and {Dietrich}, J.~P. and {Donath}, A. and {Droettboom}, M. and {Earl}, N. and {Erben}, T. and {Fabbro}, S. and {Ferreira}, L.~A. and {Finethy}, T. and {Fox}, R.~T. and {Garrison}, L.~H. and {Gibbons}, S.~L.~J. and {Goldstein}, D.~A. and {Gommers}, R. and {Greco}, J.~P. and {Greenfield}, P. and {Groener}, A.~M. and {Grollier}, F. and {Hagen}, A. and {Hirst}, P. and {Homeier}, D. and {Horton}, A.~J. and {Hosseinzadeh}, G. and {Hu}, L. and {Hunkeler}, J.~S. and {Ivezi{\'c}}, {\v{Z}}. and {Jain}, A. and {Jenness}, T. and {Kanarek}, G. and {Kendrew}, S. and {Kern}, N.~S. and {Kerzendorf}, W.~E. and {Khvalko}, A. and {King}, J. and {Kirkby}, D. and {Kulkarni}, A.~M. and {Kumar}, A. and {Lee}, A. and {Lenz}, D. and {Littlefair}, S.~P. and {Ma}, Z. and {Macleod}, D.~M. and {Mastropietro}, M. and {McCully}, C. and {Montagnac}, S. and {Morris}, B.~M. and {Mueller}, M. and {Mumford}, S.~J. and {Muna}, D. and {Murphy}, N.~A. and {Nelson}, S. and {Nguyen}, G.~H. and {Ninan}, J.~P. and {N{\"o}the}, M. and {Ogaz}, S. and {Oh}, S. and {Parejko}, J.~K. and {Parley}, N. and {Pascual}, S. and {Patil}, R. and {Patil}, A.~A. and {Plunkett}, A.~L. and {Prochaska}, J.~X. and {Rastogi}, T. and {Reddy Janga}, V. and {Sabater}, J. and {Sakurikar}, P. and {Seifert}, M. and {Sherbert}, L.~E. and {Sherwood-Taylor}, H. and {Shih}, A.~Y. and {Sick}, J. and {Silbiger}, M.~T. and {Singanamalla}, S. and {Singer}, L.~P. and {Sladen}, P.~H. and {Sooley}, K.~A. and {Sornarajah}, S. and {Streicher}, O. and {Teuben}, P. and {Thomas}, S.~W. and {Tremblay}, G.~R. and {Turner}, J.~E.~H. and {Terr{\'o}n}, V. and {van Kerkwijk}, M.~H. and {de la Vega}, A. and {Watkins}, L.~L. and {Weaver}, B.~A. and {Whitmore}, J.~B. and {Woillez}, J. and {Zabalza}, V. and {Astropy Contributors}},
        title = "{The Astropy Project: Building an Open-science Project and Status of the v2.0 Core Package}",
      journal = {\aj},
         year = 2018,
        month = sep,
       volume = {156},
       number = {3},
          eid = {123},
        pages = {123},
          doi = {10.3847/1538-3881/aabc4f},
archivePrefix = {arXiv},
       eprint = {1801.02634},
 primaryClass = {astro-ph.IM},
       adsurl = {https://ui.adsabs.harvard.edu/abs/2018AJ....156..123A}
}

@ARTICLE{Astropy2013,
       author = {{Astropy Collaboration} and {Robitaille}, Thomas P. and
         {Tollerud}, Erik J. and {Greenfield}, Perry and {Droettboom}, Michael and
         {Bray}, Erik and {Aldcroft}, Tom and {Davis}, Matt and
         {Ginsburg}, Adam and {Price-Whelan}, Adrian M. and
         {Kerzendorf}, Wolfgang E. and {Conley}, Alexander and {Crighton}, Neil and
         {Barbary}, Kyle and {Muna}, Demitri and {Ferguson}, Henry and
         {Grollier}, Fr{\'e}d{\'e}ric and {Parikh}, Madhura M. and
         {Nair}, Prasanth H. and {Unther}, Hans M. and {Deil}, Christoph and
         {Woillez}, Julien and {Conseil}, Simon and {Kramer}, Roban and
         {Turner}, James E.~H. and {Singer}, Leo and {Fox}, Ryan and
         {Weaver}, Benjamin A. and {Zabalza}, Victor and {Edwards}, Zachary I. and
         {Azalee Bostroem}, K. and {Burke}, D.~J. and {Casey}, Andrew R. and
         {Crawford}, Steven M. and {Dencheva}, Nadia and {Ely}, Justin and
         {Jenness}, Tim and {Labrie}, Kathleen and {Lim}, Pey Lian and
         {Pierfederici}, Francesco and {Pontzen}, Andrew and {Ptak}, Andy and
         {Refsdal}, Brian and {Servillat}, Mathieu and {Streicher}, Ole},
        title = "{Astropy: A community Python package for astronomy}",
      journal = {\aap},
         year = "2013",
        month = "Oct",
       volume = {558},
          eid = {A33},
        pages = {A33},
          doi = {10.1051/0004-6361/201322068},
archivePrefix = {arXiv},
       eprint = {1307.6212},
 primaryClass = {astro-ph.IM},
       adsurl = {https://ui.adsabs.harvard.edu/abs/2013A&A...558A..33A}
}

@ARTICLE{Hunter2007,
       author = {{Hunter}, John D.},
        title = "{Matplotlib: A 2D Graphics Environment}",
      journal = {Computing in Science and Engineering},
         year = 2007,
        month = may,
       volume = {9},
       number = {3},
        pages = {90-95},
          doi = {10.1109/MCSE.2007.55},
       adsurl = {https://ui.adsabs.harvard.edu/abs/2007CSE.....9...90H}
}

@ARTICLE{harris2020,
       author = {{Harris}, Charles R. and {Millman}, K. Jarrod and {van der Walt}, St{\'e}fan J. and {Gommers}, Ralf and {Virtanen}, Pauli and {Cournapeau}, David and {Wieser}, Eric and {Taylor}, Julian and {Berg}, Sebastian and {Smith}, Nathaniel J. and {Kern}, Robert and {Picus}, Matti and {Hoyer}, Stephan and {van Kerkwijk}, Marten H. and {Brett}, Matthew and {Haldane}, Allan and {del R{\'\i}o}, Jaime Fern{\'a}ndez and {Wiebe}, Mark and {Peterson}, Pearu and {G{\'e}rard-Marchant}, Pierre and {Sheppard}, Kevin and {Reddy}, Tyler and {Weckesser}, Warren and {Abbasi}, Hameer and {Gohlke}, Christoph and {Oliphant}, Travis E.},
        title = "{Array programming with NumPy}",
      journal = {\nat},
         year = 2020,
        month = sep,
       volume = {585},
       number = {7825},
        pages = {357-362},
          doi = {10.1038/s41586-020-2649-2},
archivePrefix = {arXiv},
       eprint = {2006.10256},
 primaryClass = {cs.MS},
       adsurl = {https://ui.adsabs.harvard.edu/abs/2020Natur.585..357H}
}

@ARTICLE{vanderwalt2011,
       author = {{van der Walt}, St{\'e}fan and {Colbert}, S. Chris and {Varoquaux}, Ga{\"e}l},
        title = "{The NumPy Array: A Structure for Efficient Numerical Computation}",
      journal = {Computing in Science and Engineering},
         year = 2011,
        month = mar,
       volume = {13},
       number = {2},
        pages = {22-30},
          doi = {10.1109/MCSE.2011.37},
archivePrefix = {arXiv},
       eprint = {1102.1523},
 primaryClass = {cs.MS},
       adsurl = {https://ui.adsabs.harvard.edu/abs/2011CSE....13b..22V}
}

@ARTICLE{Hasinger:2008:AGN_properties,
       author = {{Hasinger}, G.},
        title = "{Absorption properties and evolution of active galactic nuclei}",
      journal = {\aap},
         year = 2008,
        month = nov,
       volume = {490},
       number = {3},
        pages = {905-922},
          doi = {10.1051/0004-6361:200809839},
archivePrefix = {arXiv},
       eprint = {0808.0260},
 primaryClass = {astro-ph},
       adsurl = {https://ui.adsabs.harvard.edu/abs/2008A&A...490..905H}
}

@INPROCEEDINGS{Kaier:2004:PanSTARRS,
       author = {{Kaiser}, Nicholas},
        title = "{Pan-STARRS: a wide-field optical survey telescope array}",
    booktitle = {Ground-based Telescopes},
         year = 2004,
       editor = {{Oschmann}, Jr., Jacobus M.},
       series = {Society of Photo-Optical Instrumentation Engineers (SPIE) Conference Series},
       volume = {5489},
        month = oct,
        pages = {11-22},
          doi = {10.1117/12.552472},
       adsurl = {https://ui.adsabs.harvard.edu/abs/2004SPIE.5489...11K}
}

@ARTICLE{York:2000:SDSS,
       author = {{York}, Donald G. and {Adelman}, J. and {Anderson}, Jr., John E. and {Anderson}, Scott F. and {Annis}, James and {Bahcall}, Neta A. and {Bakken}, J.~A. and {Barkhouser}, Robert and {Bastian}, Steven and {Berman}, Eileen and {Boroski}, William N. and {Bracker}, Steve and {Briegel}, Charlie and {Briggs}, John W. and {Brinkmann}, J. and {Brunner}, Robert and {Burles}, Scott and {Carey}, Larry and {Carr}, Michael A. and {Castander}, Francisco J. and {Chen}, Bing and {Colestock}, Patrick L. and {Connolly}, A.~J. and {Crocker}, J.~H. and {Csabai}, Istv{\'a}n and {Czarapata}, Paul C. and {Davis}, John Eric and {Doi}, Mamoru and {Dombeck}, Tom and {Eisenstein}, Daniel and {Ellman}, Nancy and {Elms}, Brian R. and {Evans}, Michael L. and {Fan}, Xiaohui and {Federwitz}, Glenn R. and {Fiscelli}, Larry and {Friedman}, Scott and {Frieman}, Joshua A. and {Fukugita}, Masataka and {Gillespie}, Bruce and {Gunn}, James E. and {Gurbani}, Vijay K. and {de Haas}, Ernst and {Haldeman}, Merle and {Harris}, Frederick H. and {Hayes}, J. and {Heckman}, Timothy M. and {Hennessy}, G.~S. and {Hindsley}, Robert B. and {Holm}, Scott and {Holmgren}, Donald J. and {Huang}, Chi-hao and {Hull}, Charles and {Husby}, Don and {Ichikawa}, Shin-Ichi and {Ichikawa}, Takashi and {Ivezi{\'c}}, {\v{Z}}eljko and {Kent}, Stephen and {Kim}, Rita S.~J. and {Kinney}, E. and {Klaene}, Mark and {Kleinman}, A.~N. and {Kleinman}, S. and {Knapp}, G.~R. and {Korienek}, John and {Kron}, Richard G. and {Kunszt}, Peter Z. and {Lamb}, D.~Q. and {Lee}, B. and {Leger}, R. French and {Limmongkol}, Siriluk and {Lindenmeyer}, Carl and {Long}, Daniel C. and {Loomis}, Craig and {Loveday}, Jon and {Lucinio}, Rich and {Lupton}, Robert H. and {MacKinnon}, Bryan and {Mannery}, Edward J. and {Mantsch}, P.~M. and {Margon}, Bruce and {McGehee}, Peregrine and {McKay}, Timothy A. and {Meiksin}, Avery and {Merelli}, Aronne and {Monet}, David G. and {Munn}, Jeffrey A. and {Narayanan}, Vijay K. and {Nash}, Thomas and {Neilsen}, Eric and {Neswold}, Rich and {Newberg}, Heidi Jo and {Nichol}, R.~C. and {Nicinski}, Tom and {Nonino}, Mario and {Okada}, Norio and {Okamura}, Sadanori and {Ostriker}, Jeremiah P. and {Owen}, Russell and {Pauls}, A. George and {Peoples}, John and {Peterson}, R.~L. and {Petravick}, Donald and {Pier}, Jeffrey R. and {Pope}, Adrian and {Pordes}, Ruth and {Prosapio}, Angela and {Rechenmacher}, Ron and {Quinn}, Thomas R. and {Richards}, Gordon T. and {Richmond}, Michael W. and {Rivetta}, Claudio H. and {Rockosi}, Constance M. and {Ruthmansdorfer}, Kurt and {Sandford}, Dale and {Schlegel}, David J. and {Schneider}, Donald P. and {Sekiguchi}, Maki and {Sergey}, Gary and {Shimasaku}, Kazuhiro and {Siegmund}, Walter A. and {Smee}, Stephen and {Smith}, J. Allyn and {Snedden}, S. and {Stone}, R. and {Stoughton}, Chris and {Strauss}, Michael A. and {Stubbs}, Christopher and {SubbaRao}, Mark and {Szalay}, Alexander S. and {Szapudi}, Istvan and {Szokoly}, Gyula P. and {Thakar}, Anirudda R. and {Tremonti}, Christy and {Tucker}, Douglas L. and {Uomoto}, Alan and {Vanden Berk}, Dan and {Vogeley}, Michael S. and {Waddell}, Patrick and {Wang}, Shu-i. and {Watanabe}, Masaru and {Weinberg}, David H. and {Yanny}, Brian and {Yasuda}, Naoki and {SDSS Collaboration}},
        title = "{The Sloan Digital Sky Survey: Technical Summary}",
      journal = {\aj},
         year = 2000,
        month = sep,
       volume = {120},
       number = {3},
        pages = {1579-1587},
          doi = {10.1086/301513},
archivePrefix = {arXiv},
       eprint = {astro-ph/0006396},
 primaryClass = {astro-ph},
       adsurl = {https://ui.adsabs.harvard.edu/abs/2000AJ....120.1579Y}
}

@ARTICLE{Gaia:2023:DR3,
       author = {{Gaia Collaboration} and {Vallenari}, A. and {Brown}, A.~G.~A. and {Prusti}, T. and {de Bruijne}, J.~H.~J. and {Arenou}, F. and {Babusiaux}, C. and {Biermann}, M. and {Creevey}, O.~L. and {Ducourant}, C. and {Evans}, D.~W. and {Eyer}, L. and {Guerra}, R. and {Hutton}, A. and {Jordi}, C. and {Klioner}, S.~A. and {Lammers}, U.~L. and {Lindegren}, L. and {Luri}, X. and {Mignard}, F. and {Panem}, C. and {Pourbaix}, D. and {Randich}, S. and {Sartoretti}, P. and {Soubiran}, C. and {Tanga}, P. and {Walton}, N.~A. and {Bailer-Jones}, C.~A.~L. and {Bastian}, U. and {Drimmel}, R. and {Jansen}, F. and {Katz}, D. and {Lattanzi}, M.~G. and {van Leeuwen}, F. and {Bakker}, J. and {Cacciari}, C. and {Casta{\~n}eda}, J. and {De Angeli}, F. and {Fabricius}, C. and {Fouesneau}, M. and {Fr{\'e}mat}, Y. and {Galluccio}, L. and {Guerrier}, A. and {Heiter}, U. and {Masana}, E. and {Messineo}, R. and {Mowlavi}, N. and {Nicolas}, C. and {Nienartowicz}, K. and {Pailler}, F. and {Panuzzo}, P. and {Riclet}, F. and {Roux}, W. and {Seabroke}, G.~M. and {Sordo}, R. and {Th{\'e}venin}, F. and {Gracia-Abril}, G. and {Portell}, J. and {Teyssier}, D. and {Altmann}, M. and {Andrae}, R. and {Audard}, M. and {Bellas-Velidis}, I. and {Benson}, K. and {Berthier}, J. and {Blomme}, R. and {Burgess}, P.~W. and {Busonero}, D. and {Busso}, G. and {C{\'a}novas}, H. and {Carry}, B. and {Cellino}, A. and {Cheek}, N. and {Clementini}, G. and {Damerdji}, Y. and {Davidson}, M. and {de Teodoro}, P. and {Nu{\~n}ez Campos}, M. and {Delchambre}, L. and {Dell'Oro}, A. and {Esquej}, P. and {Fern{\'a}ndez-Hern{\'a}ndez}, J. and {Fraile}, E. and {Garabato}, D. and {Garc{\'\i}a-Lario}, P. and {Gosset}, E. and {Haigron}, R. and {Halbwachs}, J. -L. and {Hambly}, N.~C. and {Harrison}, D.~L. and {Hern{\'a}ndez}, J. and {Hestroffer}, D. and {Hodgkin}, S.~T. and {Holl}, B. and {Jan{\ss}en}, K. and {Jevardat de Fombelle}, G. and {Jordan}, S. and {Krone-Martins}, A. and {Lanzafame}, A.~C. and {L{\"o}ffler}, W. and {Marchal}, O. and {Marrese}, P.~M. and {Moitinho}, A. and {Muinonen}, K. and {Osborne}, P. and {Pancino}, E. and {Pauwels}, T. and {Recio-Blanco}, A. and {Reyl{\'e}}, C. and {Riello}, M. and {Rimoldini}, L. and {Roegiers}, T. and {Rybizki}, J. and {Sarro}, L.~M. and {Siopis}, C. and {Smith}, M. and {Sozzetti}, A. and {Utrilla}, E. and {van Leeuwen}, M. and {Abbas}, U. and {{\'A}brah{\'a}m}, P. and {Abreu Aramburu}, A. and {Aerts}, C. and {Aguado}, J.~J. and {Ajaj}, M. and {Aldea-Montero}, F. and {Altavilla}, G. and {{\'A}lvarez}, M.~A. and {Alves}, J. and {Anders}, F. and {Anderson}, R.~I. and {Anglada Varela}, E. and {Antoja}, T. and {Baines}, D. and {Baker}, S.~G. and {Balaguer-N{\'u}{\~n}ez}, L. and {Balbinot}, E. and {Balog}, Z. and {Barache}, C. and {Barbato}, D. and {Barros}, M. and {Barstow}, M.~A. and {Bartolom{\'e}}, S. and {Bassilana}, J. -L. and {Bauchet}, N. and {Becciani}, U. and {Bellazzini}, M. and {Berihuete}, A. and {Bernet}, M. and {Bertone}, S. and {Bianchi}, L. and {Binnenfeld}, A. and {Blanco-Cuaresma}, S. and {Blazere}, A. and {Boch}, T. and {Bombrun}, A. and {Bossini}, D. and {Bouquillon}, S. and {Bragaglia}, A. and {Bramante}, L. and {Breedt}, E. and {Bressan}, A. and {Brouillet}, N. and {Brugaletta}, E. and {Bucciarelli}, B. and {Burlacu}, A. and {Butkevich}, A.~G. and {Buzzi}, R. and {Caffau}, E. and {Cancelliere}, R. and {Cantat-Gaudin}, T. and {Carballo}, R. and {Carlucci}, T. and {Carnerero}, M.~I. and {Carrasco}, J.~M. and {Casamiquela}, L. and {Castellani}, M. and {Castro-Ginard}, A. and {Chaoul}, L. and {Charlot}, P. and {Chemin}, L. and {Chiaramida}, V. and {Chiavassa}, A. and {Chornay}, N. and {Comoretto}, G. and {Contursi}, G. and {Cooper}, W.~J. and {Cornez}, T. and {Cowell}, S. and {Crifo}, F. and {Cropper}, M. and {Crosta}, M. and {Crowley}, C. and {Dafonte}, C. and {Dapergolas}, A. and {David}, M. and {David}, P. and {de Laverny}, P. and {De Luise}, F. and {De March}, R.},
        title = "{Gaia Data Release 3. Summary of the content and survey properties}",
      journal = {\aap},
         year = 2023,
        month = jun,
       volume = {674},
          eid = {A1},
        pages = {A1},
          doi = {10.1051/0004-6361/202243940},
archivePrefix = {arXiv},
       eprint = {2208.00211},
 primaryClass = {astro-ph.GA},
       adsurl = {https://ui.adsabs.harvard.edu/abs/2023A&A...674A...1G}
}

@ARTICLE{Wenger:2000:SIMBAD,
       author = {{Wenger}, M. and {Ochsenbein}, F. and {Egret}, D. and {Dubois}, P. and {Bonnarel}, F. and {Borde}, S. and {Genova}, F. and {Jasniewicz}, G. and {Lalo{\"e}}, S. and {Lesteven}, S. and {Monier}, R.},
        title = "{The SIMBAD astronomical database. The CDS reference database for astronomical objects}",
      journal = {\aaps},
         year = 2000,
        month = apr,
       volume = {143},
        pages = {9-22},
          doi = {10.1051/aas:2000332},
archivePrefix = {arXiv},
       eprint = {astro-ph/0002110},
 primaryClass = {astro-ph},
       adsurl = {https://ui.adsabs.harvard.edu/abs/2000A&AS..143....9W}
}

@ARTICLE{Ginsburg:2019:Astroquery,
   author = {{Ginsburg}, A. and {Sip{\H o}cz}, B.~M. and {Brasseur}, C.~E. and
	{Cowperthwaite}, P.~S. and {Craig}, M.~W. and {Deil}, C. and
	{Guillochon}, J. and {Guzman}, G. and {Liedtke}, S. and {Lian Lim}, P. and
	{Lockhart}, K.~E. and {Mommert}, M. and {Morris}, B.~M. and
	{Norman}, H. and {Parikh}, M. and {Persson}, M.~V. and {Robitaille}, T.~P. and
	{Segovia}, J.-C. and {Singer}, L.~P. and {Tollerud}, E.~J. and
	{de Val-Borro}, M. and {Valtchanov}, I. and {Woillez}, J. and
	{The Astroquery collaboration} and {a subset of the astropy collaboration}
	},
    title = "{astroquery: An Astronomical Web-querying Package in Python}",
  journal = {\aj},
archivePrefix = "arXiv",
   eprint = {1901.04520},
 primaryClass = "astro-ph.IM",
     year = 2019,
    month = mar,
   volume = 157,
      eid = {98},
    pages = {98},
      doi = {10.3847/1538-3881/aafc33},
   adsurl = {https://adsabs.harvard.edu/abs/2019AJ....157...98G}
}

@ARTICLE{Wainer:2025:47Tuc,
       author = {{Wainer}, Tobin M. and {Davenport}, James R.~A. and {Bellm}, Eric C. and {Yuankun} and {Wang} and {Caplar}, Neven and {Burdett}, Elliott S. and {Shipp}, Nora and {Parejko}, John K. and {Thoron}, Gray and {Butler}, Eric and {Salwa}, Maya and {Howard}, Erin Leigh and {Smart}, Brianna Marie and {Beebe}, Wilson and {Ghosh-Coutinho}, Ishan F. and {Abel}, Bob and {Ivezi{\'c}}, {\v{Z}}eljko},
        title = "{Crowded Field Photometry with Rubin: Exploring 47 Tucanae with Data Preview 1}",
      journal = {arXiv e-prints},
         year = 2025,
        month = jul,
          eid = {arXiv:2507.03228},
        pages = {arXiv:2507.03228},
          doi = {10.48550/arXiv.2507.03228},
archivePrefix = {arXiv},
       eprint = {2507.03228},
 primaryClass = {astro-ph.GA},
       adsurl = {https://ui.adsabs.harvard.edu/abs/2025arXiv250703228W}
}

@ARTICLE{Choi:2025:47TucDP1,
       author = {{Choi}, Yumi and {Olsen}, Knut A.~G. and {Carlin}, Jeffrey L. and {Yuankun} and {Wang} and {Moolekamp}, Fred and {Saha}, Abi and {Sullivan}, Ian and {Slater}, Colin T. and {Tucker}, Douglas L. and {Adair}, Christina L. and {Ferguson}, Peter S. and {Kang}, Yijung and {Pe{\~n}a Ram{\'\i}rez}, Karla and {Rabus}, Markus},
        title = "{47 Tuc in Rubin Data Preview 1: Exploring Early LSST Data and Science Potential}",
      journal = {arXiv e-prints},
         year = 2025,
        month = jul,
          eid = {arXiv:2507.01343},
        pages = {arXiv:2507.01343},
          doi = {10.48550/arXiv.2507.01343},
archivePrefix = {arXiv},
       eprint = {2507.01343},
 primaryClass = {astro-ph.SR},
       adsurl = {https://ui.adsabs.harvard.edu/abs/2025arXiv250701343C}
}

@ARTICLE{Malanchev:2025:LSDB_DP1,
       author = {{Malanchev}, Konstantin and {DeLucchi}, Melissa and {Caplar}, Neven and {Malz}, Alex I. and {Beebe}, Wilson and {Branton}, Doug and {Campos}, Sandro and {Connolly}, Andrew and {Dai}, Mi and {Kubica}, Jeremy and {Lynn}, Olivia and {Mandelbaum}, Rachel and {McGuire}, Sean and {Aubourg}, Eric and {Blum}, Robert David and {Carlin}, Jeffrey L. and {Delgado}, Francisco and {Gangler}, Emmanuel and {Jannuzi}, Buell T. and {Jenness}, Tim and {Kang}, Yijung and {Kannawadi}, Arun and {Moniez}, Marc and {Plazas Malag{\'o}n}, Andr{\'e}s A. and {van Reeven}, Wouter and {Sanmartim}, David and {Urbach}, Elana K. and {Wood-Vasey}, W.~M.},
        title = "{Variability-finding in Rubin Data Preview 1 with LSDB}",
      journal = {arXiv e-prints},
         year = 2025,
        month = jun,
          eid = {arXiv:2506.23955},
        pages = {arXiv:2506.23955},
          doi = {10.48550/arXiv.2506.23955},
archivePrefix = {arXiv},
       eprint = {2506.23955},
 primaryClass = {astro-ph.IM},
       adsurl = {https://ui.adsabs.harvard.edu/abs/2025arXiv250623955M}
}

@article{Budavari:08:BayesianCrossmatch,
	author = {{Budav{\'a}ri}, T. and {Szalay}, A.~S.},
	journal = {\apj},
	month = may,
	pages = {301-309},
	title = {{Probabilistic Cross-Identification of Astronomical Sources}},
	volume = 679,
	year = 2008}

@article{Shi:19:ILPCrossmatch,
	author = {{Shi}, Xiaochen and {Budav{\'a}ri}, Tam{\'a}s and {Basu}, Amitabh},
	journal = {\apj},
	month = {Jan},
	number = {1},
	pages = {51},
	title = {{Probabilistic Cross-identification of Multiple Catalogs in Crowded Fields}},
	volume = {870},
	year = {2019}}

@article{Budavari:16:Probabilistic-C,
	author = {{Budav{\'a}ri}, T. and {Basu}, A.},
	journal = {\aj},
	month = oct,
	pages = {86},
	title = {{Probabilistic Cross-identification in Crowded Fields as an Assignment Problem}},
	volume = 152,
	year = 2016}

@article{Haakonsen:09:XIDII,
	author = {{Haakonsen}, C.~B. and {Rutledge}, R.~E.},
	journal = {\apjs},
	month = sep,
	pages = {138-151},
	title = {{XID II: Statistical Cross-Association of ROSAT Bright Source Catalog X-ray Sources with 2MASS Point Source Catalog Near-Infrared Sources}},
	volume = 184,
	year = 2009}

@article{Rutledge:00:RASSXID,
	author = {{Rutledge}, R.~E. and {Brunner}, R.~J. and {Prince}, T.~A. and {Lonsdale}, C.},
	journal = {\apjs},
	month = nov,
	pages = {335-353},
	title = {{XID: Cross-Association of ROSAT/Bright Source Catalog X-Ray Sources with USNO A-2 Optical Point Sources}},
	volume = 131,
	year = 2000}

@misc{Gwyn:2024:gaia_lsst_bands,
  author       = {{Gwyn}, Stephen},
  title        = "Mapping Gaia magnitudes to LSST bands",
  howpublished = "LSST Community Forum",
  month        = jul,
  year         = 2024,
  url         = {https://community.lsst.org/t/mapping-gaia-magnitudes-to-lsst-bands/8927/2}
}

@misc{object,
    author = "{NSF-DOE Vera C. Rubin Observatory}",
    doi = "10.71929/RUBIN/2570324",
    url = "https://www.osti.gov//servlets/purl/2570324",
    title = "{Legacy Survey of Space and Time Data Preview 1: object dataset type [Data set]}",
    publisher = "NSF-DOE Vera C. Rubin Observatory",
    year = "2025"
}

@misc{dia_object,
    author = "{NSF-DOE Vera C. Rubin Observatory}",
    doi = "10.71929/RUBIN/2570318",
    url = "https://www.osti.gov//servlets/purl/2570318",
    title = "{Legacy Survey of Space and Time Data Preview 1: dia\_object dataset type [Data set]}",
    publisher = "NSF-DOE Vera C. Rubin Observatory",
    year = "2025"
}

@misc{dia_source,
    author = "{NSF-DOE Vera C. Rubin Observatory}",
    doi = "10.71929/RUBIN/2570316",
    url = "https://www.osti.gov//servlets/purl/2570316",
    title = "{Legacy Survey of Space and Time Data Preview 1: dia\_source dataset type [Data set]}",
    publisher = "NSF-DOE Vera C. Rubin Observatory",
    year = "2025"
}

@misc{deepcoadd,
  doi = {10.71929/RUBIN/2570313},
  url = {https://www.osti.gov//servlets/purl/2570313},
  author = {{NSF-DOE Vera C. Rubin Observatory}},
  title = {Legacy Survey of Space and Time Data Preview 1: deep\_coadd dataset type},
  publisher = {NSF-DOE Vera C. Rubin Observatory},
  year = {2025}
}

@misc{visit_image,
    author = "{NSF-DOE Vera C. Rubin Observatory}",
    doi = "10.71929/RUBIN/2570311",
    url = "https://www.osti.gov//servlets/purl/2570311",
    title = "{Legacy Survey of Space and Time Data Preview 1: visit\_image dataset type [Data set]}",
    publisher = "NSF-DOE Vera C. Rubin Observatory",
    year = "2025"
}

@misc{difference_image,
    author = "{NSF-DOE Vera C. Rubin Observatory}",
    doi = "10.71929/RUBIN/2570312",
    url = "https://www.osti.gov//servlets/purl/2570312",
    title = "{Legacy Survey of Space and Time Data Preview 1: difference\_image dataset type [Data set]}",
    publisher = "NSF-DOE Vera C. Rubin Observatory",
    year = "2025"
}

@misc{ForcedSourceOnDiaObject,
    author = "{NSF-DOE Vera C. Rubin Observatory}",
    doi = "10.71929/RUBIN/2570320",
    url = "https://www.osti.gov//servlets/purl/2570320",
    title = "{Legacy Survey of Space and Time Data Preview 1: dia\_object\_forced\_source dataset type [Data set]}",
    publisher = "NSF-DOE Vera C. Rubin Observatory",
    year = "2025"
}

@misc{ForcedSourceOnObject,
    author = "{NSF-DOE Vera C. Rubin Observatory}",
    doi = "10.71929/RUBIN/2570326",
    url = "https://www.osti.gov//servlets/purl/2570326",
    title = "{Legacy Survey of Space and Time Data Preview 1: object\_forced\_source dataset type [Data set]}",
    publisher = "NSF-DOE Vera C. Rubin Observatory",
    year = "2025"
}

@ARTICLE{Drake:2014:CRTS_CVs,
       author = {{Drake}, A.~J. and {G{\"a}nsicke}, B.~T. and {Djorgovski}, S.~G. and {Wils}, P. and {Mahabal}, A.~A. and {Graham}, M.~J. and {Yang}, T. -C. and {Williams}, R. and {Catelan}, M. and {Prieto}, J.~L. and {Donalek}, C. and {Larson}, S. and {Christensen}, E.},
        title = "{Cataclysmic variables from the Catalina Real-time Transient Survey}",
      journal = {\mnras},
         year = 2014,
        month = jun,
       volume = {441},
       number = {2},
        pages = {1186-1200},
          doi = {10.1093/mnras/stu639},
archivePrefix = {arXiv},
       eprint = {1404.3732},
 primaryClass = {astro-ph.SR},
       adsurl = {https://ui.adsabs.harvard.edu/abs/2014MNRAS.441.1186D}
}

@ARTICLE{Kato:2017:SU_UMa_CVs,
       author = {{Kato}, Taichi and {Isogai}, Keisuke and {Hambsch}, Franz-Josef and {Vanmunster}, Tonny and {Itoh}, Hiroshi and {Monard}, Berto and {Tordai}, Tam{\'a}s and {Kimura}, Mariko and {Wakamatsu}, Yasuyuki and {Kiyota}, Seiichiro and {Miller}, Ian and {Starr}, Peter and {Kasai}, Kiyoshi and {Shugarov}, Sergey Yu. and {Chochol}, Drahomir and {Katysheva}, Natalia and {Zaostrojnykh}, Anna M. and {Seker{\'a}{\v{s}}}, Matej and {Kuznyetsova}, Yuliana G. and {Kalinicheva}, Eugenia S. and {Golysheva}, Polina and {Krushevska}, Viktoriia and {Maeda}, Yutaka and {Dubovsky}, Pavol A. and {Kudzej}, Igor and {Pavlenko}, Elena P. and {Antonyuk}, Kirill A. and {Pit}, Nikolaj V. and {Sosnovskij}, Aleksei A. and {Antonyuk}, Oksana I. and {Baklanov}, Aleksei V. and {Pickard}, Roger D. and {Kojiguchi}, Naoto and {Sugiura}, Yuki and {Tei}, Shihei and {Yamamura}, Kenta and {Matsumoto}, Katsura and {Ruiz}, Javier and {Stone}, Geoff and {Cook}, Lewis M. and {de Miguel}, Enrique and {Akazawa}, Hidehiko and {Goff}, William N. and {Morelle}, Etienne and {Kafka}, Stella and {Littlefield}, Colin and {Bolt}, Greg and {Dubois}, Franky and {Brincat}, Stephen M. and {Maehara}, Hiroyuki and {Sakanoi}, Takeshi and {Kagitani}, Masato and {Imada}, Akira and {Voloshina}, Irina B. and {Andreev}, Maksim V. and {Sabo}, Richard and {Richmond}, Michael and {Rodda}, Tony and {Nelson}, Peter and {Nazarov}, Sergey and {Mishevskiy}, Nikolay and {Myers}, Gordon and {Denisenko}, Denis and {Stanek}, Krzysztof Z. and {Shields}, Joseph V. and {Kochanek}, Christopher S. and {Holoien}, Thomas W. -S. and {Shappee}, Benjamin and {Prieto}, Jos{\'e} L. and {Itagaki}, Koh-ichi and {Nishiyama}, Koichi and {Kabashima}, Fujio and {Stubbings}, Rod and {Schmeer}, Patrick and {Muyllaert}, Eddy and {Horie}, Tsuneo and {Shears}, Jeremy and {Poyner}, Gary and {Moriyama}, Masayuki},
        title = "{Survey of period variations of superhumps in SU UMa-type dwarf novae. IX. The ninth year (2016-2017)}",
      journal = {\pasj},
         year = 2017,
        month = oct,
       volume = {69},
       number = {5},
          eid = {75},
        pages = {75},
          doi = {10.1093/pasj/psx058},
archivePrefix = {arXiv},
       eprint = {1706.03870},
 primaryClass = {astro-ph.SR},
       adsurl = {https://ui.adsabs.harvard.edu/abs/2017PASJ...69...75K}
}

@ARTICLE{diClemente:1996:AGN_SF,
       author = {{di Clemente}, A. and {Giallongo}, E. and {Natali}, G. and {Trevese}, D. and {Vagnetti}, F.},
        title = "{The Variability of Quasars. II. Frequency Dependence}",
      journal = {\apj},
         year = 1996,
        month = jun,
       volume = {463},
        pages = {466},
          doi = {10.1086/177261},
archivePrefix = {arXiv},
       eprint = {astro-ph/9512159},
 primaryClass = {astro-ph},
       adsurl = {https://ui.adsabs.harvard.edu/abs/1996ApJ...463..466D}
}

@ARTICLE{DeCicco:2022:AGN_SF_VLT,
       author = {{De Cicco}, D. and {Bauer}, F.~E. and {Paolillo}, M. and {S{\'a}nchez-S{\'a}ez}, P. and {Brandt}, W.~N. and {Vagnetti}, F. and {Pignata}, G. and {Radovich}, M. and {Vaccari}, M.},
        title = "{A structure function analysis of VST-COSMOS AGN}",
      journal = {\aap},
         year = 2022,
        month = aug,
       volume = {664},
          eid = {A117},
        pages = {A117},
          doi = {10.1051/0004-6361/202142750},
archivePrefix = {arXiv},
       eprint = {2205.12275},
 primaryClass = {astro-ph.GA},
       adsurl = {https://ui.adsabs.harvard.edu/abs/2022A&A...664A.117D}
}

@ARTICLE{Fagin:2025:AGN_sims,
       author = {{Fagin}, Joshua and {Chan}, James Hung-Hsu and {Best}, Henry and {O'Dowd}, Matthew and {Ford}, K.~E. Saavik and {Graham}, Matthew J. and {Park}, Ji Won and {Villar}, V. Ashley},
        title = "{Joint Modeling of Quasar Variability and Accretion Disk Reprocessing Using Latent Stochastic Differential Equations}",
      journal = {\apj},
         year = 2025,
        month = jul,
       volume = {988},
       number = {1},
          eid = {59},
        pages = {59},
          doi = {10.3847/1538-4357/addabc},
archivePrefix = {arXiv},
       eprint = {2410.18423},
 primaryClass = {astro-ph.GA},
       adsurl = {https://ui.adsabs.harvard.edu/abs/2025ApJ...988...59F}
}

@ARTICLE{Roster:2025:Euclid_Xray,
       author = {{Euclid Collaboration} and {Roster}, W. and {Salvato}, M. and {Buchner}, J. and {Shirley}, R. and {Lusso}, E. and {Landt}, H. and {Zamorani}, G. and {Siudek}, M. and {Laloux}, B. and {Matamoro Zatarain}, T. and {Ricci}, F. and {Fotopoulou}, S. and {Ferr{\'e}-Mateu}, A. and {Lopez Lopez}, X. and {Aghanim}, N. and {Altieri}, B. and {Amara}, A. and {Andreon}, S. and {Auricchio}, N. and {Aussel}, H. and {Baccigalupi}, C. and {Baldi}, M. and {Balestra}, A. and {Bardelli}, S. and {Battaglia}, P. and {Biviano}, A. and {Bonchi}, A. and {Branchini}, E. and {Brescia}, M. and {Brinchmann}, J. and {Camera}, S. and {Ca{\~n}as-Herrera}, G. and {Capobianco}, V. and {Carbone}, C. and {Carretero}, J. and {Casas}, S. and {Castellano}, M. and {Castignani}, G. and {Cavuoti}, S. and {Chambers}, K.~C. and {Cimatti}, A. and {Colodro-Conde}, C. and {Congedo}, G. and {Conselice}, C.~J. and {Conversi}, L. and {Copin}, Y. and {Courbin}, F. and {Courtois}, H.~M. and {Cropper}, M. and {Da Silva}, A. and {Degaudenzi}, H. and {De Lucia}, G. and {Di Giorgio}, A.~M. and {Dolding}, C. and {Dole}, H. and {Dubath}, F. and {Duncan}, C.~A.~J. and {Dupac}, X. and {Dusini}, S. and {Escoffier}, S. and {Fabricius}, M. and {Farina}, M. and {Farinelli}, R. and {Faustini}, F. and {Ferriol}, S. and {Finelli}, F. and {Fosalba}, P. and {Fourmanoit}, N. and {Frailis}, M. and {Franceschi}, E. and {Galeotta}, S. and {George}, K. and {Gillis}, B. and {Giocoli}, C. and {Gracia-Carpio}, J. and {Granett}, B.~R. and {Grazian}, A. and {Grupp}, F. and {Gwyn}, S. and {Haugan}, S.~V.~H. and {Holmes}, W. and {Hook}, I.~M. and {Hormuth}, F. and {Hornstrup}, A. and {Hudelot}, P. and {Jahnke}, K. and {Jhabvala}, M. and {Keih{\"a}nen}, E. and {Kermiche}, S. and {Kiessling}, A. and {Kubik}, B. and {K{\"u}mmel}, M. and {Kunz}, M. and {Kurki-Suonio}, H. and {Le Boulc'h}, Q. and {Le Brun}, A.~M.~C. and {Le Mignant}, D. and {Ligori}, S. and {Lilje}, P.~B. and {Lindholm}, V. and {Lloro}, I. and {Mainetti}, G. and {Maino}, D. and {Maiorano}, E. and {Mansutti}, O. and {Marcin}, S. and {Marggraf}, O. and {Martinelli}, M. and {Martinet}, N. and {Marulli}, F. and {Massey}, R. and {Masters}, D.~C. and {Medinaceli}, E. and {Mei}, S. and {Melchior}, M. and {Mellier}, Y. and {Meneghetti}, M. and {Merlin}, E. and {Meylan}, G. and {Mora}, A. and {Moresco}, M. and {Moscardini}, L. and {Nakajima}, R. and {Neissner}, C. and {Niemi}, S. -M. and {Nightingale}, J.~W. and {Padilla}, C. and {Paltani}, S. and {Pasian}, F. and {Pedersen}, K. and {Percival}, W.~J. and {Pettorino}, V. and {Pires}, S. and {Polenta}, G. and {Poncet}, M. and {Popa}, L.~A. and {Pozzetti}, L. and {Raison}, F. and {Rebolo}, R. and {Renzi}, A. and {Rhodes}, J. and {Riccio}, G. and {Romelli}, E. and {Roncarelli}, M. and {Saglia}, R. and {Sakr}, Z. and {S{\'a}nchez}, A.~G. and {Sapone}, D. and {Sartoris}, B. and {Schewtschenko}, J.~A. and {Schirmer}, M. and {Schneider}, P. and {Schrabback}, T. and {Secroun}, A. and {Seidel}, G. and {Seiffert}, M. and {Serrano}, S. and {Simon}, P. and {Sirignano}, C. and {Sirri}, G. and {Stanco}, L. and {Steinwagner}, J. and {Tallada-Cresp{\'\i}}, P. and {Tavagnacco}, D. and {Taylor}, A.~N. and {Tereno}, I. and {Toft}, S. and {Toledo-Moreo}, R. and {Torradeflot}, F. and {Tutusaus}, I. and {Valenziano}, L. and {Valiviita}, J. and {Vassallo}, T. and {Verdoes Kleijn}, G. and {Veropalumbo}, A. and {Wang}, Y. and {Weller}, J. and {Zacchei}, A. and {Zerbi}, F.~M. and {Zinchenko}, I.~A. and {Zucca}, E. and {Allevato}, V. and {Ballardini}, M. and {Bolzonella}, M. and {Bozzo}, E. and {Burigana}, C. and {Cabanac}, R. and {Cappi}, A. and {Di Ferdinando}, D. and {Escartin Vigo}, J.~A. and {Gabarra}, L. and {Huertas-Company}, M. and {Mart{\'\i}n-Fleitas}, J. and {Matthew}, S. and {Mauri}, N. and {Metcalf}, R.~B. and {Pezzotta}, A. and {P{\"o}ntinen}, M. and {Porciani}, C.},
        title = "{Euclid Quick Data Release (Q1). Optical and near-infrared identification and classification of point-like X-ray selected sources}",
      journal = {arXiv e-prints},
         year = 2025,
        month = mar,
          eid = {arXiv:2503.15316},
        pages = {arXiv:2503.15316},
          doi = {10.48550/arXiv.2503.15316},
archivePrefix = {arXiv},
       eprint = {2503.15316},
 primaryClass = {astro-ph.GA},
       adsurl = {https://ui.adsabs.harvard.edu/abs/2025arXiv250315316E}
}

@INPROCEEDINGS{Reuter:2016:LSST_opsim,
       author = {{Reuter}, Michael A. and {Cook}, Kem H. and {Delgado}, Francisco and {Petry}, Catherine E. and {Ridgway}, Stephen T.},
        title = "{Simulating the LSST OCS for conducting survey simulations using the LSST scheduler}",
    booktitle = {Modeling, Systems Engineering, and Project Management for Astronomy VI},
         year = 2016,
       editor = {{Angeli}, George Z. and {Dierickx}, Philippe},
       series = {Society of Photo-Optical Instrumentation Engineers (SPIE) Conference Series},
       volume = {9911},
        month = aug,
          eid = {991125},
        pages = {991125},
          doi = {10.1117/12.2232680},
       adsurl = {https://ui.adsabs.harvard.edu/abs/2016SPIE.9911E..25R}
}

@software{wang_2026_18665437,
  author       = {Wang, Yuankun},
  title        = {ykwang1/DP1\_Xray\_public: Optical Counterparts to
                   X-ray sources in LSST DP1 (rev. 1)
                  },
  month        = feb,
  year         = 2026,
  publisher    = {Zenodo},
  version      = {v1.1},
  doi          = {10.5281/zenodo.18665437},
  url          = {https://doi.org/10.5281/zenodo.18665437},
  swhid        = {swh:1:dir:f6e8bb240848827d382fe7c1793bcefc8f672a16
                   ;origin=https://doi.org/10.5281/zenodo.18665389;vi
                   sit=swh:1:snp:5f1c684bf56b4949e2b66f84b5c1a0139a12
                   cf07;anchor=swh:1:rel:2868e2891e5c6dfea4c8788320a9
                   0ee30a1d1e6c;path=ykwang1-DP1\_Xray\_public-bba8ece
                  },
}

\end{document}